\documentclass[twocolumn,aps,prl] {revtex4} 
\usepackage{graphicx}
\begin{document}
\pagestyle{empty} 
\title{Contact mechanics for randomly rough surfaces}
\author{B.N.J. Persson}
\affiliation{IFF, FZ-J\"ulich, 52425 J\"ulich, Germany}

\begin{abstract}
When two solids are squeezed together they will in general not make atomic
contact everywhere within the nominal (or apparent) contact area. This fact
has huge practical implications and must be considered in many technological
applications. In this paper I briefly review basic theories of contact mechanics.
I consider in detail a
recently developed contact mechanics theory. 
I derive 
boundary conditions for the stress probability distribution function  
for elastic, elastoplastic and adhesive contact between solids and
present numerical results illustrating some aspects of the theory.
I analyze contact problems for very smooth polymer
(PMMA) and Pyrex glass surfaces prepared by
cooling liquids of glassy materials from above the glass transition temperature.
I show that the surface roughness which results from the frozen capillary waves 
can have a large influence on the contact between the solids. 
The analysis suggest a new explanation for 
puzzling experimental results [L. Bureau, T. Baumberger and C. Caroli,
arXiv:cond-mat/0510232] about the dependence
of the frictional shear stress on the load for contact between a glassy polymer 
lens and flat substrates. 
I discuss the possibility of testing the theory using numerical
methods, e.g., finite element calculations.
\end{abstract}
\maketitle


{\bf 1 Introduction}

{\bf 2 Surface roughness}

{ 2.1 Road surfaces}

{ 2.2 Surfaces with frozen capillary waves}

{\bf 3 Contact mechanics theories}

{ 3.1 Hertz theory}

{ 3.2 Greenwood-Williamson theory}

{ 3.3 The theory of Bush, Gibson and Thomas}

{ 3.4 Persson theory}

{ 3.5 Comparison with numerical results}

{ 3.6 Macroasperity contact}

{ 3.7 On the validity of contact mechanics theories}

{\bf 4 Elastic contact mechanics}

{\bf 5 Elastic contact mechanics with graded elasticity}

{\bf 6 Elastoplastic contact mechanics: constant hardness}

{\bf 7 Elastoplastic contact mechanics: size-dependent hardness}

{\bf 8 Elastic contact mechanics with adhesion}

{ 8.1 Boundary condition and physical interpretation}

{ 8.2 Elastic energy}

{ 8.3 Numerical results}

{ 8.4 Detachment stress}

{ 8.5 Flaw tolerant adhesion}

{ 8.6 Contact mechanics and adhesion for viscoelastic solids}

{ 8.7 Strong adhesion for (elastically) stiff materials: Natures solution}

{\bf 9 Elastoplastic contact mechanics with adhesion}

{\bf 10 Applications}

{ 10.1 Contact mechanics for PMMA}
 
{ 10.2 Contact mechanics for Pyrex glass}
 
{\bf 11 On the philosophy of contact mechanics}

{\bf 12 Comments on numerical studies of contact mechanics}

{\bf 13 Summary and conclusion}

{\bf Appendix A: Power spectrum of frozen capillary waves}

{\bf Appendix B: Stress distribution in adhesive contact between smooth spherical surfaces}

\vskip 0.5cm

{\bf 1 Introduction}

\begin{figure}
\includegraphics[width=0.35\textwidth]{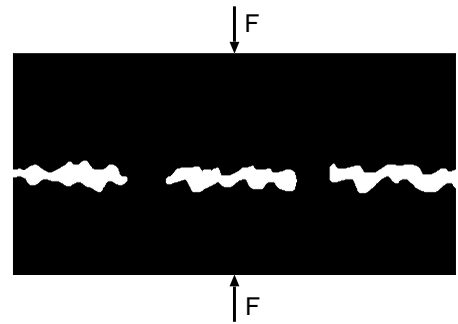} 
\caption{\label{contact}
Two solid blocks squeezed in contact with the force $F$. 
The area of real (atomic) contact $A$ is usually
an extremely small fraction of the nominal (or apparent) contact area $A_0$.}
\end{figure}

How large is the area of real contact when two solids are 
brought into contact (fig. \ref{contact})? This fundamental
question has interested scientists since the pioneering work of Hertz published 1882\cite{Hertz1}.
The problem is of tremendous practical importance, since the area of real contact
influence a large number of physical properties such as the contact resistivity,
heat transfer, adhesion, friction and the wear between solids in stationary or sliding contact.
The contact area and the gap between two solids is also of 
crucial important for seals and for the optical properties of composite
systems, e.g., the optical interference
between glass lenses. 
 
Hertz studied the frictionless contact between elastic solids with smooth
surface profiles which could be approximated as parabolic close to the contact area.
This theory predict that the contact area $A$ increases non-linearly with the squeezing force
$F_{\rm N}$ as $A\sim F_{\rm N}^{2/3}$. 
The simplest model of a randomly rough but nominal flat surface consist of a regular array 
of spherical bumps (or cups) with equal radius of curvature $R$ and equal height. 
If such a surface is squeezed against an elastic solid with a flat surface, the Hertz
contact theory can be (approximately) applied to each asperity contact region and one expect 
that the area of real contact should scale as 
$F_{\rm N}^{2/3}$ with the load or squeezing force $F_{\rm N}$. However, this is not in accordance with
experiment for randomly rough surfaces, 
which shows that, even when the contact is purely elastic (i.e., no plastic flow), 
the area of real contact is proportional 
to $F_{\rm N}$ as long as the contact area $A$ is small compared to the nominal
contact area $A_0$. This is, for example, the physical origin of Coulombs friction law which states that
the friction force is proportional to the normal load\cite{BookP}. 

In a pioneering study, Archard\cite{Archard} showed that in a more realistic model of rough surfaces, where
the roughness is described by a hierarchical model consisting of small spherical bumps on top 
of larger spherical bums and so on, the area of real contact is proportional to the load. 
This model explain the basic physics in a clear manner, but it cannot be used in practical calculations
because real surfaces cannot be represented by the idealized surface roughness assumed by Archard.
A somewhat more useful model from the point of view of applications, was presented by
Greenwood and Williamson\cite{GreenW}. They described the rough surface as consisting of spherical bumps of
equal radius of curvature $R$, but with a Gaussian distribution of heights. This model predict
that the area of real contact is {\it nearly} proportional to the load. A more refined model
based on the same picture was developed by Bush et al\cite{Bush}. They approximated the summits with
paraboloids to which they applied the Hertzian contact theory. The height 
distribution was described by a random
process, and they found that at low squeezing force the area of 
real contact increases linearly with $F_{\rm N}$. Several other contact theories are reviewed in
Ref. \cite{GZ}.

All the contact mechanical models described above assumes that the area of real contact $A$ 
is very small
compared to the nominal contact area $A_0$, and 
made up from many 
small Hertzian-like asperity contact regions. 
This formulation has two drawbacks. First it neglects the
interaction between the different contact regions. That is, if an asperity is squeezed against a 
flat hard surface it will deform not just locally, but the elastic deformation field will
extend some distance away from the asperity and hence 
influence the deformation of other asperities\cite{P66}.
Secondly, the contact mechanics theories can only be applied as long as $A<<A_0$.

A fundamentally different approach to contact mechanics has recently been developed by
Persson\cite{P8,P1}. It starts from the 
opposite limit of very large contact, and is in fact exact in the limit of
complete contact (that is, the pressure distribution at the interface is exact). 
It also accounts (in an approximate way) for the elastic coupling between asperity contact regions. 
For small squeezing force it predict that the contact area $A$ is proportional to the
load $F_{\rm N}$, while as $F_{\rm N}$ increases $A$ approach $A_0$ 
in a continuous manner. 
Thus, the theory is (approximately) valid for {\it all} squeezing forces. 
In addition, the theory
is very flexible and can be applied to more complex situations such as contact mechanics involving 
viscoelastic solids\cite{P8,Cre}, adhesion\cite{P1} and plastic yield\cite{P8}. 
It was originally developed in the context of rubber friction
on rough surfaces (e.g., road surfaces)\cite{P8}, where the (velocity dependent) 
viscoelastic deformations of the rubber on many different length scales are 
accurately described by the theory.
The only input related to the surface roughness is the surface roughness power spectrum $C(q)$, which
can be easily obtained from the surface height profile $h({\bf x})$ (see Ref. \cite{P3}).
The height profile can be measured routinely
on all relevant length scales
using different methods such as the Atomic Force 
Microscope (AFM), and optical methods\cite{P3}.

In this paper I present a brief review of contact mechanics.
I also present new results and a few comments
related to a recently developed contact mechanics theory. I derive 
boundary conditions for the stress probability distribution function  
for elastic, elastoplastic and adhesive contact between solids and
present numerical results illustrating some aspects of the theory.
I analyze contact problems for very smooth polymer
(PMMA) and Pyrex glass surfaces prepared by
cooling liquids of glassy materials from above the glass transition temperature.
I show that the surface roughness which results from the frozen capillary waves 
can have a large influence on the contact between the solids. 
I also discuss the possibility of testing the theory using numerical
methods, e.g., finite element calculations.

\vskip 0.5 cm
{\bf 2 Surface roughness}
 
The most important property of a rough surface is the surface roughness power spectrum
which is the Fourier transform of the height-height correlation function\cite{Nayak}:
$$C({\bf q})={1\over (2\pi )^2}\int d^2x \ \langle h({\bf x})h({\bf 0})\rangle e^{-i{\bf q}\cdot {\bf x}}$$
Here $z=h({\bf x})$ is the height of the surface at the point ${\bf x}=(x,y)$ above 
a flat reference plane chosen so that $\langle h({\bf x})\rangle = 0$. The angular
bracket $\langle ... \rangle$ stands for ensemble averaging. In what follows we consider only surfaces 
for which the statistical properties are isotropic so that $C({\bf q})=C(q)$ only depend on the
magnitude $|{\bf q}|$ of the wavevector ${\bf q}$. 

Many surfaces of interest are approximately self affine fractal. A fractal surface
has the property that if one magnify a section of it, it looks the same.
For a self affine fractal surface, the statistical properties 
of the surface are unchanged if we make a scale
change, which in general is different 
for the perpendicular direction as compared to
the parallel (in plane) directions.
The power spectrum of a self affine fractal surface
$$C(q)\sim q^{-2(H+1)}$$
where the Hurst exponent $H$ is related to the fractal dimension via $D_{\rm f} = 3-H$. In reality,
surfaces cannot be self affine fractal over all length scales. Thus, the largest wave vector
possible is $q_1 \approx 2 \pi /a$ where $a$ is some atomic distance or lattice constant,
and the smallest possible wave vector is $q_L \approx 2 \pi /L$, where $L$ is the linear size of
the surface. Surfaces prepared by fracture of brittle materials may be self affine fractal in the whole
length scale regime from $a$ to $L$, but most surfaces are (approximately)
self affine fractal only in some finite wave vector regime, 
or not self affine fractal anywhere as for, e.g., 
surfaces prepared by slowly cooling a glassy material below the glass transition temperature, see below.

Randomly rough surfaces with any given power spectrum $C(q)$ can be generated using\cite{P3}
$$h({\bf x}) = \sum_{\bf q} B({\bf q}) e^{i[{\bf q}\cdot {\bf x}+\phi({\bf q})]}\eqno(1)$$
where $\phi ({\bf q})$ are independent random variables, uniformly distributed in the
interval $[0,2\pi [$, and with $B({\bf q})=(2\pi/L)[C({\bf q})]^{1/2}$, where $L=A_0^{1/2}$. 

\begin{figure}
\includegraphics[width=0.35\textwidth]{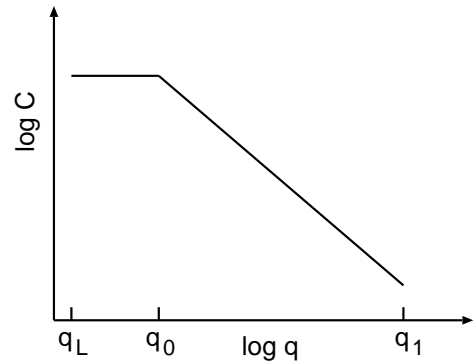} 
\caption{\label{Cq1}
Surface roughness power spectrum of a surface which is self affine fractal for
$q_1>q>q_0$. The long-distance roll-off wave vector $q_0$ and the short distance cut-off
wave vector $q_1$ depend on the system under consideration. The slope of
the ${\rm log} C-{\rm log}q$ relation for $q > q_0$ determines the fractal
exponent of the surface. The lateral size $L$ of the surface (or of the studied surface region)
determines the smallest possible wave vector $q_L=2\pi /L$.}
\end{figure}

Technologically important surfaces can be prepared by blasting 
small hard particles (e.g., sand) against
a smooth substrate. Such surfaces are typically (approximately) self affine 
fractal in some wave vector regime $q_0 < q <q_1$
as illustrated in Fig. \ref{Cq1}. The roll-off wave vector $q_0$ decreases with increasing time of blasting.
Sand blasted surfaces, and surfaces prepared by crack propagation, have typical
fractal dimension $D_{\rm f} \approx 2.2$, corresponding to the slope 
$-2(H+1) = -3.6$ 
of the ${\rm log} C-{\rm log}q$ relation for $q > q_0$. 
We now give two example of the power spectrum of rough surfaces.
 
\vskip 0.3cm
{\bf 2.1 Road surfaces}

\begin{figure}
\includegraphics[width=0.45\textwidth]{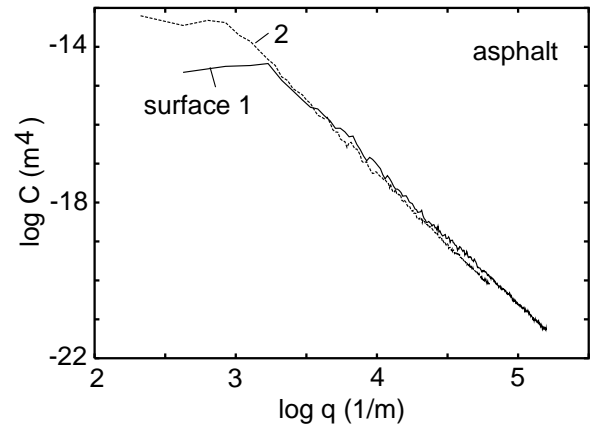}
\caption{\label{CqOpel}
The
surface roughness power spectra $C(q)$ for two asphalt road surfaces.}
\end{figure}

\begin{figure}
\includegraphics[width=0.45\textwidth]{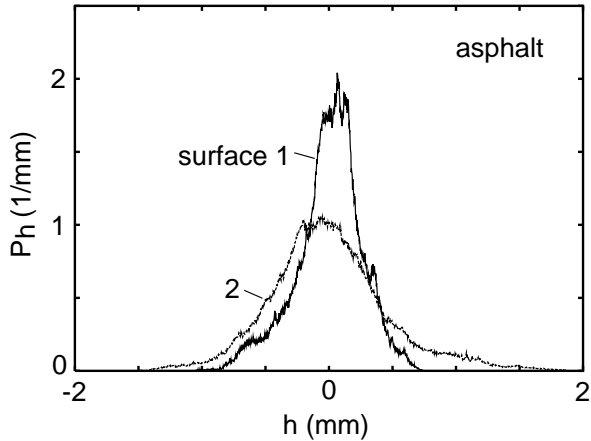}
\caption{\label{asphaltPh}
The height distribution $P_h$ for two different asphalt road surfaces.
}
\end{figure}

Asphalt and concrete road pavements have nearly perfect self-affine
fractal power spectra, with a very well-defined roll-off wave vector $q_0
= 2\pi /\lambda_0$ of order $1000 \ {\rm m}^{-1}$, corresponding to
$\lambda_0 \approx 1 \ {\rm cm}$, which reflects the largest stone
particles used in the asphalt. This is illustrated in Fig.~\ref{CqOpel}
for two different asphalt pavements. From the slope of the curves for
$q>q_0$ one can deduce the fractal dimension $D_{\rm f} \approx 2.2$,
which is typical for asphalt and concrete road surfaces. The height
distributions of the two asphalt surfaces are shown in
Fig.~\ref{asphaltPh}.
Note that the {\it rms} roughness amplitude of surface {\bf 2} is nearly
twice as high as for surface ${\bf 1}$.  Nevertheless, the tire-rubber
friction\cite{P8} is slightly higher on the road surface {\bf 1} because it has
slightly larger power spectra for most $q$-values in Fig.~\ref{CqOpel}.
Thus there is in general {\it no} direct correlation between the {\it
rms}-roughness amplitude and the rubber friction on road surfaces.

\vskip 0.3cm
{\bf 2.2 Surfaces with frozen capillary waves}

Many technological applications require surfaces of solids to be extremely
smooth. For example, window glass has to be so smooth that no (or negligible)
diffusive scattering of the light occur (a glass surface with strong 
roughness on the length scale
of the light wavelength will appear white and non-transparent because
of diffusive light scattering). For glassy materials, e.g., 
silicate glasses, or glassy polymers, e.g., Plexiglas, 
extremely flat surfaces can be prepared by
cooling the liquid 
from well above the glass transition temperature $T_{\rm g}$, since in the liquid state 
the surface tension tends to eliminate (or reduce)
short-wavelength roughness. Thus,
float glass (e.g., window glass) is produced by letting melted glass flow into
a bath of molten tin in a continuous ribbon. The glass
and the tin do not mix and the contact surface between these
two materials is (nearly) perfectly flat. Cooling of the melt below the glass transition
temperature produces a solid glass with extremely flat surfaces.
Similarly, spherical particles with 
very smooth surfaces can be prepared by cooling liquid drops of glassy materials 
below the glass
transition temperature. Sometimes glassy objects are ``fire polished'' 
by exposing them to a flame
or an intense laser beam\cite{Lag} which melt a 
thin surface layer of the material, which 
will flow and form a very smooth surface in order to minimize
the surface free energy. In this way, small-wavelength roughness is
reduced while the overall (macroscopic) shape of the solid object is unchanged.  

\begin{figure}
  \includegraphics[width=0.45\textwidth]{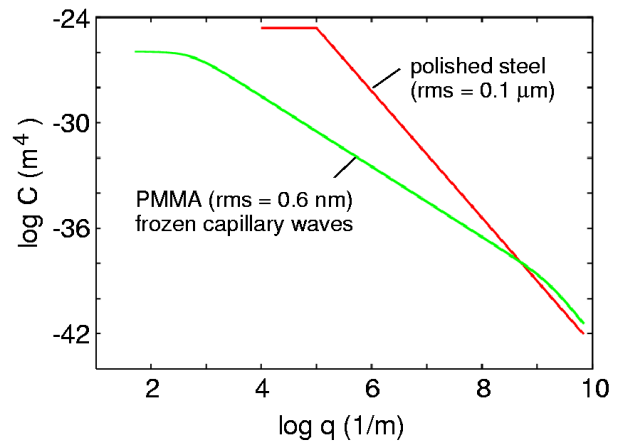}
  \caption{ \label{C.frozen}
The surface roughness power spectrum $C(q)$ for a polished steel surface with
the root-mean-square (rms) roughness $0.1 \ {\rm \mu m}$ and the fractal dimension
$D_{\rm f}=2.2$, and for a PMMA surface prepared by cooling the melted polymer below the
glass transition temperature. In this case the rms roughness is $0.6 \ {\rm nm}$. 
}
\end{figure}

However, surfaces of glassy materials prepared by cooling the liquid 
from a temperature above $T_{\rm g}$ cannot be perfectly
(molecularly) smooth, but exhibit a fundamental minimum surface roughness,
with a maximum height fluctuation  amplitude of typically $10 \ {\rm nm}$, 
which cannot be eliminated by any changes in the cooling procedure.
The reason is that at the surface of a liquid, fluctuations of vertical displacement are caused by
thermally excited capillary waves (ripplons)\cite{Jack,Die}. At the surface of very viscous 
supercooled liquids near the glass transition temperature these fluctuations
become very slow and are finally frozen in at the glass transition\cite{Jack}. 

The roughness derived from the frozen capillary waves is unimportant in many practical
applications, e.g., in most optical applications as is vividly evident for glass windows. 
However, in other applications they may
be of profound importance. Here we will show that they are of crucial importance in
contact mechanics. For elastically hard solids such as silica glass, already a
roughness of order $1 \ {\rm nm}$ is enough to eliminate the adhesion, and reduce the contact area 
between two
such surfaces to just a small fraction of the nominal 
contact area even for high squeezing pressures\cite{P3}.
This is the case even for elastically much more compliant solids such
as Plexiglas (PMMA), see Sec. 10.1. 

The surface roughness power spectrum due to capillary waves is of the form
(see Appendix A)\cite{Die,Buzza,Sey}
$$C(q) = {1\over (2 \pi )^2} {k_BT\over \rho g +\gamma q^2 +\kappa q^4}\eqno(2)$$
where $\gamma$ is the surface tension, $\kappa$ the bending stiffness,
and $\rho$ the mass density of the glassy melt.
In Fig. \ref{C.frozen} we compare the 
the surface roughness power spectrum $C(q)$ for a (typical) highly polished steel surface with
the root-mean-square (rms) roughness $0.1 \ {\rm \mu m}$ and the fractal dimension
$D_{\rm f}=2.2$, with that of a PMMA surface prepared by cooling the melted polymer below the
glass transition temperature. In this case the rms roughness is $0.6 \ {\rm nm}$. 
Note that in spite of the much larger rms roughness of the polished steel surface, for very
large wavevectors the power spectrum of the PMMA surface is larger. 
This fact is of extreme importance for contact mechanics, since the large $q$-region
usually dominates the contact mechanics. 

In Sec. 10 we will study contact mechanics problems for PMMA and for Pyrex glass.
The smallest relevant wavevector $q = q_L \approx 2\pi /L$, where $L$ is the diameter 
of the nominal contact area $A_0$. In the applications below $L\approx 100 \ {\rm \mu m}$
(for PMMA) and $L \approx 1 {\rm \mu m}$ (for Pyrex)  giving $\gamma q^2 >> \rho g$.
Thus, the gravity term in (2) can be neglected. 
The mean of the square of the surface height fluctuation is given by
$$\langle h^2 \rangle = \int d^2q \ C(q) = 2\pi \int_{q_L}^{q_1} dq \ qC(q)$$
$$= {k_BT \over 2 \pi \gamma} {\rm ln} \left [ {q_1\over q_L} 
\left ({q_L^2+q_c^2 \over q_1^2+q_c^2}\right )^{1/2}\right ]
\approx {k_BT \over 2 \pi \gamma} {\rm ln} \left ( {q_c\over q_L}\right )
\eqno(3)$$
where $q_c = (\gamma/\kappa)^{1/2}$ is a cross-over wavevector.
Eq. (3) gives the 
rms roughness $\approx 0.5 \ {\rm nm}$ for PMMA and $\approx 0.15 \ {\rm nm}$ for Pyrex,
which typically correspond
to maximum roughness amplitude 
of about $5 \ {\rm nm}$ and $1.5 \ {\rm nm}$, respectively (see Appendix A). 

\vskip 0.5 cm
{\bf 3 Contact mechanics theories}

We first briefly review the Hertz contact theory for elastic spheres with perfectly smooth
surfaces. 
Most contact mechanics theories for randomly rough surfaces approximate the surface asperities
with spherical or elliptical ``bumps'' to which they apply the Hertz contact theory. 
In most cases the elastic coupling between the asperity contact regions is neglected
which is a good approximation only if the (average) distance between 
nearby contact regions is large enough. 
A necessary (but not sufficient) condition for this is that the normal (squeezing) 
force must be so small that 
the area of real contact is very small
compared to the nominal contact area.
This is the basic
strategy both for the Greenwood-Williamson theory and the (more accurate) theory of Bush et al, and we
describe briefly both these theories. 
The theory of Persson start from the opposite limit of complete contact, and the theory 
is (approximately) valid for all squeezing forces. 

\begin{figure}[htb]
\includegraphics[width=0.2\textwidth]{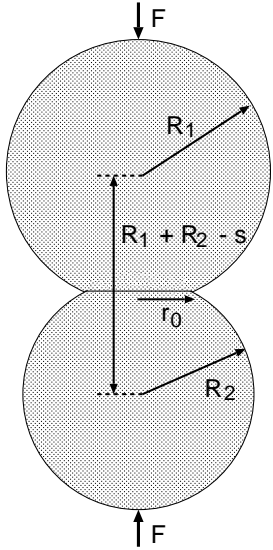}
\caption{
Contact between two elastic spherical objects. The spheres are squeezed together with
the force $F$ and a circular contact area (radius $r_0$) is formed.
} 
\label{ball1}
\end{figure}

\vskip 0.3cm
{\bf 3.1 Hertz theory}

The Hertz theory describes the contact between two elastic spherical bodies 
(radius $R_1$ and $R_2$, respectively) with perfectly
smooth surfaces. Assume that the spheres are squeezed together by the force $F$. 
The deformation field in the solids can be determined by minimizing the elastic
deformation energy. 
The radius $r_0$ of the circular contact region (see Fig. \ref{ball1}) is given by
$$r_0 = \left ({R_1R_2\over R_1+R_2}\right )^{1/3} \left ({3F (1-\nu^2) \over 4E}
\right )^{1/3}\eqno(4)$$
where
$${1-\nu^2 \over E} =   
{1-\nu_1^2 \over E_1 }
+{1-\nu_2^2 \over E_2 }\eqno(5)$$
where $E_1$ and $E_2$ are the elastic modulus of the solids and $\nu_1$ and
$\nu_2$ the corresponding Poisson ratios. The distance $s$ the two solids approach each other
(the penetration, see Fig. \ref{ball1}) is given by
$$s=
\left ({R_1+R_2\over R_1 R_2}\right )^{1/3} 
\left ({3F (1-\nu^2) \over 4E}\right )^{2/3}\eqno(6)$$
For the special case of a sphere (radius $R$) in contact with a 
flat surface we get from (5) and (6)
the area of contact
$$\pi r_0^2 = \pi R s\eqno(7)$$
and the squeezing force
$$F= {4E\over 3(1-\nu^2)} s^{3/2}R^{1/2}.\eqno(8)$$
The pressure distribution in the contact area depends only on the distance
$r$ from the center of the circular contact area:
$$\sigma (r) = \sigma_0 \left [1-\left (r\over r_0\right )^2\right ]^{1/2}$$
where $\sigma_0 = F/\pi r_0^2$ is the average pressure.

The Hertz contact theory has been generalized to adhesive contact. In this case the deformation field
in the solids is determined by minimizing the elastic energy plus the interfacial binding energy.
In the simplest case it is assumed that the wall-wall interaction has an infinitesimal extent
so that the binding energy is of the form $\pi r_0^2 \Delta \gamma$, where $\Delta \gamma =
\gamma_1+\gamma_2-\gamma_{12}$ is the change in the surface free energy (per unit area) as 
two flat surfaces of the
solids makes contact. The resulting Johnson--Kendall--Roberts
(JKR) theory\cite{JKR} is in good agreement with experiment\cite{Is}. For solids
which are elastically very hard it is no longer possible to neglect the finite extent of the
wall-wall interaction and other contact mechanics theories must be used in this case. 
The limiting case of a rigid sphere in contact with a flat rigid substrate was considered by
Derjaguin\cite{Derj} in 1934, and result in a pull-off force $F_{\rm pull-off} = 2 \pi \Delta \gamma R$
which is a factor of $4/3$ larger as predicted by the JKR theory.
A more accurate theory, which takes into account {\it both} the elasticity of the solids and
the finite extent of the wall-wall interaction, will give a pull-off force
somewhere between the JKR and Derjaguin limits.
In Appendix B 
I discuss a particular elegant and pedagogical approach to the 
adhesive contact mechanics for spherical bodies, which has
been developed by Greenwood and Johnson, and which takes into account both the elasticity and
the finite extent of the
wall-wall interaction.

\begin{figure}[htb]
\includegraphics[width=0.4\textwidth]{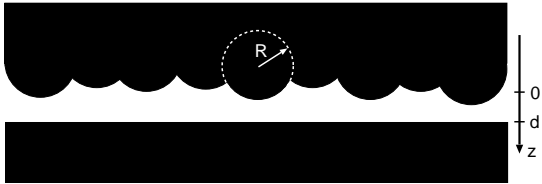}
\caption{
Contact between rough surface and a plane. All asperities with height $z$ greater than the 
separation $d$ make contact.
} 
\label{green}
\end{figure}

\vskip 0.3cm
{\bf 3.2 Greenwood-Williamson theory}

We consider the friction less contact between elastic 
solids with randomly rough surfaces. If $z=h_1({\bf x})$ and 
$h_2({\bf x})$ describes the surface height profiles, and 
$E_1$ and $E_2$ are the Young's elastic modulus of the 
two solids, and $\nu_1$ and $\nu_2$ the corresponding Poisson ratios, then the elastic 
contact problem is equivalent to the contact between a rigid solid with the roughness 
profile $h({\bf x})=h_1({\bf x})+h_2({\bf x})$, 
in contact with an elastic solid with a flat surface and with the 
Young's modulus $E$ and Poisson ratio $\nu$ chosen so that (5) is obeyed.

\begin{figure}[htb]
\includegraphics[width=0.40\textwidth]{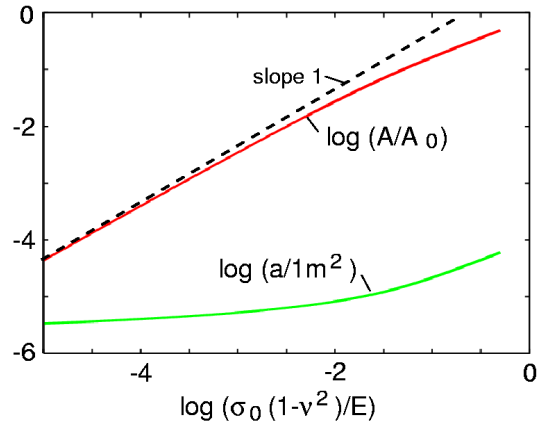}
\caption{
The logarithm of the relative contact area $A/A_0$ and the logarithm
of the average area $a$ of a contact region, as a function of the logarithm of the
nominal squeezing pressure $\sigma_0$ (in units of $E/(1-\nu^2)$). The results are
for the Greenwood-Williamson model assuming roughness on a single length scale
$\lambda_{\rm m} = 2\pi /q_{\rm m}$ with $q_{\rm m} = 600 \ {\rm m^{-1}}$
and $q_{\rm m}h_{\rm rms} = 1$. 
} 
\label{greenAa}
\end{figure}

Let us assume that there is roughness on a single length scale. We approximate the
asperities as spherical bumps with equal radius of curvature $R$ 
(see Fig. \ref{green}) and with a Gaussian 
height distribution
$$P_h = {1\over (2\pi )^{1/2} h^*} {\rm exp} \left (-{h^2 \over 2{h^*}^2} \right ),\eqno(9)$$
where $h^*$ is the root-mean-square amplitude of the summit height fluctuations,
which is slightly smaller than the root-mean-square roughness amplitude $h_{\rm rms}$ of
the surface profile $h({\bf x})$.
Assume that we can neglect the (elastic) interaction between the asperity contact regions.
If the separation between the surfaces is denoted by $d$, an asperity with height $h>d$ will
make contact with the plane and the penetration $s=h-d$.
Thus, using the Hertz contact theory with $s=h-d$ 
the (normalized) area of real contact\cite{GreenW}
$${\Delta A\over A_0} = \pi n_0 R \int_d^\infty dh \ (h-d)P_h\eqno(10)$$
where $A_0$ is the nominal contact area and
$n_0$ 
the number of asperities per unit area.
The number of contacting asperities per unit area
$${N\over A_0} = n_0 
\int_d^\infty dh \ P_h\eqno(11)$$
and the nominal squeezing stress
$$\sigma_0 = {F_{\rm N}\over A_0} = 
{4E\over 3(1-\nu^2)} n_0 \int_d^\infty dh \  (h-d)^{3/2}R^{1/2} P_h.\eqno(12)$$
In Fig. \ref{greenAa} I show, for a typical case, 
the logarithm of the relative contact area $A/A_0$ and the logarithm
of the average area $a$ of a contact patch, as a function of the logarithm of the
nominal squeezing pressure $\sigma_0$ (in units of $E/(1-\nu^2)$). Note that the contact area
depends (weakly) non-linearly on the squeezing pressure $\sigma_0$ and that the (average) contact patch 
area $a$ increases with
the squeezing pressure. Here we have defined
$$a= {\sum_i A_i^2 \over \sum_i A_i}$$
where $A_i$ is the area of contact patch $i$. Another definition, which gives
similar results, is $a=A/N= \sum_i A_i/N$.

If, as assumed above, the roughness occur on a 
single length scale, say with wavevectors with magnitudes in a narrow region
around $q_{\rm m}$ i.e., $C(q) \approx C_0 \delta (q-q_{\rm m})$, then 
we can estimate the concentration of asperities, $n_0$, the (average) radius of curvature
$R$ of the asperities, and the rms summit height fluctuation $h^*$ as follows.
We expect the asperities to form a hexagonal-like (but somewhat disordered)
distribution with the ``lattice constant'' $\lambda_{\rm m} = 2 \pi /q_{\rm m}$ so that
$$n_0 \approx 2/(\lambda_{\rm m}^2\surd 3) = {q_{\rm m}^2 \over 2 \pi^2 \surd 3} \approx 0.029 q_{\rm m}^2$$
The height profile along some axis $x$ in the surface plane will oscillate 
(``quasi-periodically'') with the wavelength
of order $\approx \lambda_{\rm m}$ roughly as 
$h(x) \approx h_1 {\rm cos}(q_{\rm m} x)$ where $h_1 = \surd 2 h_{\rm rms}$. 
Thus, 
$${1\over R} \approx h''(0) = q_{\rm m}^2 h_1 =\surd 2 q_{\rm m}^2 h_{\rm rms}.$$
Finally, we expect the rms of the fluctuation in the summit height to be somewhat
smaller than $h_0$, typically\cite{Nayak} $h^* \approx 0.6 h_{\rm rms}$.  Nayak\cite{Nayak}
has shown how to exactly relate $n_0$, $R$ and $h^*$ to $q_m$ and $h_{\rm rms}$. 

Fuller and Tabor have generalized the Greenwood-Williamson theory to the adhesive
contact between randomly rough surfaces\cite{Fuller}. 
They applied the JKR asperity contact model to each
asperity contact area, instead of the Hertz model used in the original theory.
The model presented above can also be easily generalized to elastoplastic contact. 

\vskip 0.3cm
{\bf 3.3 The theory of Bush, Gibson and Thomas}

The Greenwood-Williamson theory assumes roughness on a single length scale, which
result in an area of real contact which depends (slightly) non-linearly on the load even for very
small load. Bush et al\cite{Bush} have developed a more general and accurate contact mechanics theory where 
roughness is assumed to occur on {\it many different length scale}. This result in an area of real
contact which is proportional to the squeezing force for small squeezing force, i.e., 
$A  \sim F_{\rm N}$ as long as $A << A_0$. Thus,  
the area of real contact {\it $A$ (for small load) is strictly proportional to the
load only when roughness occur on many different length scales}.
In this case the stress distribution at the interface, and the average size
$a$ of an asperity contact region, are {\it independent} of the applied load. That is, as
the load is increased (but $A<<A_0$) existing contact areas grow and 
new contact regions form in such a way that the quantities 
referred to above remains unchanged. 

Bush et al considered the elastic contact between an isotropically
rough surface with a plane by approximating the summits of a random process model by paraboloids
with the same principle curvature and applying the Hertzian solution for their deformation.
Like in the Greenwood-Williamson theory, the elastic interaction between the asperity
contact regions is neglected. For small load the theory predict that the 
area of real contact is proportional to the load, $A=\alpha F_{\rm N}$, where
$$\alpha = \kappa {(1-\nu^2) \over E}\left ( \int d^2q \ q^2 C(q)\right )^{-1/2}\eqno(13)$$
where $\kappa = (2\pi )^{1/2}$.

\begin{figure}[htb]
   \includegraphics[width=0.4\textwidth]{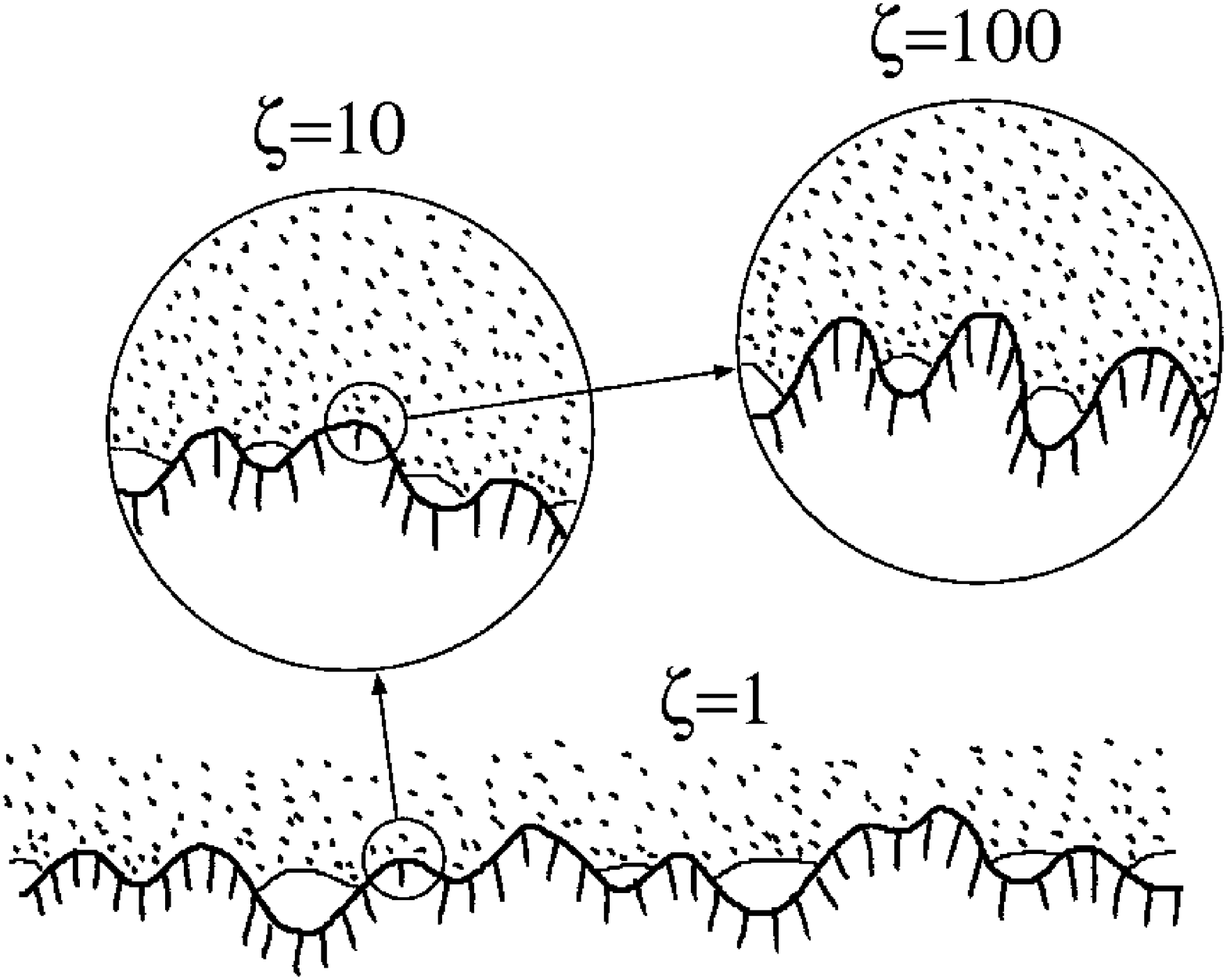}
\caption{
An elastic (e.g., rubber) block (dotted area) in adhesive contact with a hard
rough substrate (dashed area). The substrate has roughness on many different
length scales and the rubber makes partial contact with the substrate on all length scales.
When a contact area
is studied at low magnification ($\zeta=1$)
it appears as if complete contact occurs in the macro asperity contact regions,
but when the magnification is increased it is observed that in reality only partial
contact occurs.
}
\label{1x}
\end{figure}

\vskip 0.3cm
{\bf 3.4 Persson theory}

Recently, a contact mechanics theory has been developed which is valid not only
when the area of real contact is small compared to the nominal contact area, but which is
particularly accurate when the squeezing force is so high that nearly complete contact occurs
within the nominal contact area\cite{P8,P1,P66}. All other contact mechanics theories\cite{Jon,GreenW,Bush} 
were developed for the case where the area of real contact is much smaller than the
nominal contact area. The theory developed in
Ref. \cite{P8,P1} can also be applied when the adhesional interaction is included.

Fig.~\ref{1x} shows the
contact between two solids at increasing magnification $\zeta$. At low magnification
($\zeta = 1$)
it looks as if complete contact occurs between the solids at many {\it macro asperity}
contact regions,
but when the magnification is increased smaller length scale roughness is detected,
and it is observed that only partial contact occurs at the asperities.
In fact, if there would be no short distance cut-off, the true contact area
would vanish. In reality, however,
a short distance cut-off will always exist since the shortest possible length is
an atomic distance. In many cases the local pressure at asperity contact regions
at high magnification will become so high
that the material yields plastically before reaching the atomic dimension.
In these cases
the size of the real contact area will be determined mainly by the yield stress
of the solid.

The magnification $\zeta$ refers to some (arbitrary) chosen reference length scale.
This could be, e.g., the lateral size $L$ of the nominal contact area in which case 
$\zeta= L/\lambda = q/q_L$, where $\lambda$ is the shortest wavelength
roughness which can be resolved at magnification $\zeta$. Sometimes, when the
surface roughness power spectrum is of the form shown in Fig. \ref{Cq1}, we will instead
use the roll-off wavelength $\lambda_0=2\pi /q_0$ as the reference 
length so that $\zeta = \lambda_0/\lambda=q/q_0$.

Let us define the stress distribution at the magnification $\zeta$ 
$$P(\sigma, \zeta) = \langle \delta (\sigma - \sigma({\bf x},\zeta))\rangle \eqno(14)$$
Here $\sigma ({\bf x},\zeta)$ is the stress at the interface calculated when only
the surface roughness components with wave vectors $q<\zeta q_L$ is included. 
The angular brackets $\langle ... \rangle$ in (14) stands for ensemble average or,
what is in most cases equivalent, average over the surface area
$$P(\sigma, \zeta) = {1\over A_0} \int_A d^2x \ \delta (\sigma - \sigma({\bf x},\zeta))\eqno(15)$$
where $A$ is the area of contact. If the integral in (15) would be over the whole surface area
$A_0$ then $P(\sigma, \zeta)$ would have a delta function $[(A_0-A)/A_0] \delta (\sigma )$
but this term is excluded with the definition we use. The area of real contact, projected on the $xy$-plane,
can be obtained directly from the stress distribution since from (15) it follows that
$${A(\zeta)\over A_0} = \int d\sigma P(\sigma, \zeta)\eqno(16)$$
We will often denote $A(\zeta)/A_0 = P(\zeta )$.

In Ref. \cite{P8} I derived an expression for the stress distribution $P(\sigma, \zeta)$ 
at the interface 
when the interface is studied at the magnification 
$\zeta = L/\lambda$, where $L$ is the diameter of the nominal contact area 
between the solids and $\lambda$ the shortest surface roughness wavelength 
which can be detected at the resolution $\zeta$. That is, the surface roughness profile at the
magnification $\zeta$ is obtained by restricting the sum in (1) to $q < q_L \zeta$.
Assuming complete contact one can show that $P(\sigma, \zeta)$ satisfies the diffusion-like equation
$${\partial P \over \partial \zeta} = f(\zeta) {\partial^2 P \over \partial \sigma^2}\eqno(17)$$
where
$$f(\zeta)={\pi \over 4} \left ({E\over 1-\nu^2}\right )^2 q_L q^3 C(q)\eqno(18)$$
where
$q_L=2\pi /L$ and $q=\zeta q_L$. It is now assumed that (17) holds {\it locally} also when only partial
contact occur at the interface. The calculation of the stress distribution in the latter case 
involves solving (17) with appropriate boundary conditions.

If a rectangular elastic block is squeezed against the substrate with the (uniform) 
stress $\sigma_0$, then at the lowest
magnification $\zeta=1$ where the substrate appear flat, we have
$$P(\sigma,1)=\delta(\sigma-\sigma_0)\eqno(19)$$
which form an ``initial'' condition.
In addition, two boundary conditions along the $\sigma$-axis are necessary in order to
solve (17). For elastic contact $P(\sigma, \zeta)$ must vanish
as $\sigma \rightarrow \infty$. In the absence of an adhesional interaction,
the stress distribution must also vanish for $\sigma < 0$ since no tensile stress
is possible without adhesion. In fact, we will show below (Sec. 4) that $P(\sigma,\zeta)$ must 
vanish continuously as $\sigma \rightarrow 0$.
We will also derive the boundary conditions for elastoplastic contact (Sec. 6 and 7), 
and for elastic contact with adhesion (Sec. 8).  

Eq. (17) is easy to solve with the ``initial'' condition (19) and the boundary condition
$P(0,\zeta)=0$. The area of (apparent) contact when the system is studied at the magnification
$\zeta$ is given by
$${A(\zeta)\over A_0} = {1\over \surd \pi}\int_0^{\surd G} dx \ e^{-x^2/4} = 
{\rm erf}(1/2\surd G)\eqno(20)$$
where
$$G(\zeta) = {\pi \over 4} \left ( {E\over (1-\nu^2)\sigma_0}\right ) 
\int_{q_{\rm L}}^{\zeta q_{\rm L}} dq \ q^3 C(q)\eqno(21)$$
In Fig. \ref{C.frozen}(a) we show 
the dependence of the normalized contact area (at the highest magnification)
$A/A_0$ on the squeezing pressure $\sigma_0$ (in units of $E/(1-\nu^2)$). In (b)
we show how $A(\zeta)/A_0$ depend on the magnification 
(where $\zeta = 1$ now refers to 
the resolution $\lambda_0= 2\pi/q_0$ rather than the size $L$
of the surface) for $\sigma_0 (1-\nu^2)/E = 0.001$. 
The results have been obtained from (20) and (21) for the
surface ``polished steel'' with the power spectra shown 
shown in Fig. \ref{C.frozen}.

When the squeezing force $F_{\rm N}=\sigma_0 A_0$ is so small that $A<<A_0$, the equations 
above reduces to $A=\alpha F_{\rm N}$ with $\alpha$ given by (13) with
$\kappa = (8/\pi )^{1/2}$, i.e., the Persson theory also
predict that the contact area increases linearly with the squeezing force $F_{\rm N}$
as long as the contact area is small compared to the nominal contact area.

\begin{figure}[htb]
   \includegraphics[width=0.4\textwidth]{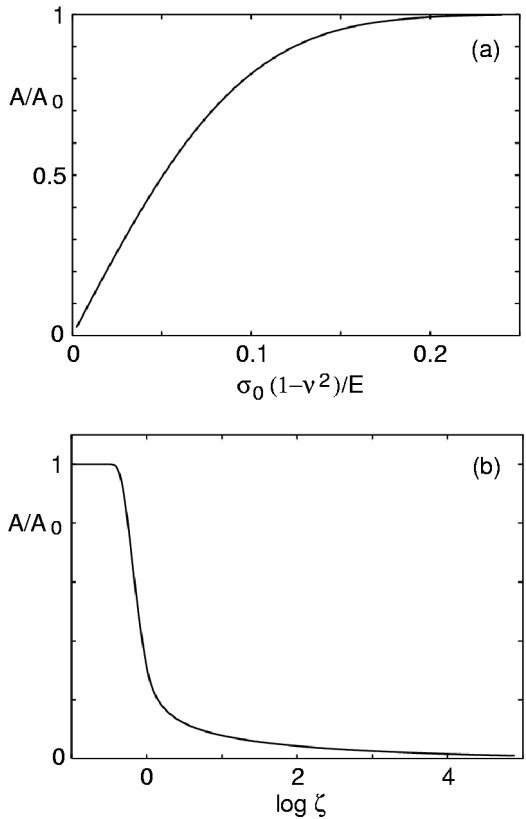}
\caption{
(a) The dependence of the normalized contact area (at the highest magnification)
$A/A_0$ on the squeezing pressure $\sigma_0$ (in units of $E/(1-\nu^2)$),
and (b) on the magnification for $\sigma_0 (1-\nu^2)/E = 0.001$. For the
surface ``polished steel'' with the power spectra shown 
shown in Fig. \ref{C.frozen}.
}
\label{1y}
\end{figure}

\vskip 0.3cm
{\bf 3.5 Comparison with numerical results}

The theory of Bush et al, and of Persson 
predict that the area of contact
increases linearly with the load for small load. In the standard theory of Greenwood and
Williamson\cite{GreenW} this result holds only approximately and a comparison of the prediction of their
theory with the present theory is therefore difficult.
The theory of Bush et al.\cite{Bush} includes roughness on many different length scales, and 
at small
load the area of contact depends linearly on the
load $A=\alpha F_{\rm N}$, where
$$\alpha =\kappa {(1-\nu^2) \over E} \left (\int d^2q \ q^2 C(q) \right )^{-1/2}\eqno(22)$$
where 
$\kappa= (2\pi )^{1/2}$.
The same expression is valid in the Persson theory but with $\kappa = (8/\pi )^{1/2}$.
Thus the latter theory predict that the contact area is a factor of $2/\pi$ smaller
than 
the theory of Bush et al. 
Both the theory of Greenwood and Williamson and that of Bush et al.,
assume that the asperity contact regions are
independent. However, as discussed in Ref.~\cite{P66},
for real surfaces (which always have
surface roughness on many different length scales) this will never be
the case even at a very low nominal contact pressure. As we argued\cite{P66} this may be the
origin of the $2/\pi$-difference between the Persson theory (which assumes roughness on many
different length scales) and the result of Bush et al.

The predictions of the contact theories of Bush et al.\cite{Bush} and Persson\cite{P8}
were compared to
numerical calculations (see Ref.~\cite{P66}\cite{summera4}).
Borri-Brunetto et al.\cite{summera2} studied the contact between self affine fractal surfaces
using an essentially exact numerical method. They found that the contact area is proportional to the squeezing
force for small squeezing forces. Furthermore, it was found that the slope $\alpha (\zeta)$
of the line $A=\alpha (\zeta) F_{\rm N}$
decreased with increasing magnification $\zeta$. This is also predicted by the analytical theory
[Eq. (22)]. In fact, good agreement was found between theory and 
computer simulations for the change in the slope with magnification and its
dependence on the fractal dimension $D_{\rm f}$.

\begin{figure}[htb]
\includegraphics[width=0.40\textwidth]{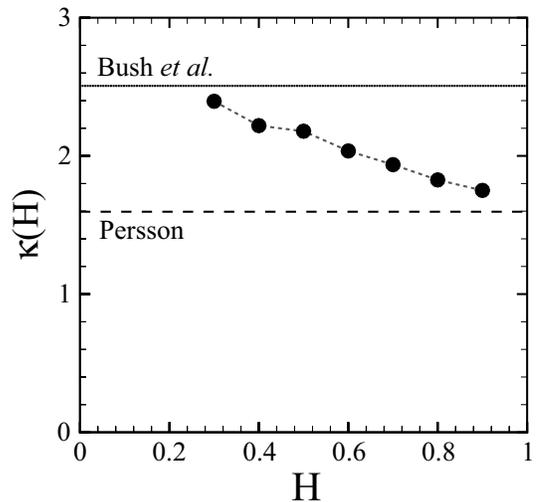}
\caption{
Dots: the factor $\kappa$ as a function of Hurst's exponent $H$ for
self affine fractal surfaces. The two horizontal lines
are the predictions of the theories of Bush et al. (solid line)
and Persson (dashed line). From Ref.~\protect\cite{summera4}.}
\label{Mark}
\end{figure}

Hyun et al. performed a finite-element analysis of contact between
elastic self-affine surfaces.
The simulations were done for a rough elastic surface contacting a perfectly rigid flat
surface.
The elastic solid was discretized into blocks and the surface nodes form a square grid.
The contact algorithm identified all nodes on the top surface that attempt to penetrate
the flat bottom surface. The total contact area $A$ was obtained by multiplying the number
of penetrating nodes by the area of each square associated with each node.
As long as the squeezing force was so small that the contact area remained below $10 \%$ 
of the nominal contact area, $A/A_0 < 0.1$, the area of real contact was found 
to be proportional to the squeezing force.
Fig.~\ref{Mark} shows Hyun et al.'s results for
the factor $\kappa$ in (22) as a function of Hurst's exponent $H$ for
self affine fractal surfaces. The two horizontal lines
are predictions of the theories of Bush et al. (solid line)
and Persson (dashed line). The agreement with the analytical predictions is quite
good considering the ambiguities in the discretization of the surface (Sec. 12). The algorithm only
consider nodal heights and assumes that contact of a node implies contact over the entire corresponding
square. This procedure would be accurate if the spacing between nodes where much
smaller than the typical size of asperity contacts.
However, the majority of the contact area consists of
clusters containing only one or a few nodes.
Since the number of large clusters grows as $H\rightarrow 1$, this may
explain why the numerical results approach Persson's prediction in this limit.

Hyun et al. also studied the distribution of connected contact regions and the contact morphology.
In addition, the interfacial stress distribution was considered, and it was found that the stress
distribution remained non-zero as the stress $\sigma \rightarrow 0$.
This violates the boundary condition
that $P(\sigma, \zeta)=0$ for $\sigma = 0$. However,
it has been shown analytically\cite{P66} that for
``smooth'' surface roughness this latter condition
must be satisfied, and the
violation of this boundary condition in the numerical
simulations reflect the way the solid was
discretized and the way the contact area was 
defined in Hyun et al.'s numerical procedure (see also Sec. 12).

In Ref. \cite {YTP}
the
constant $\alpha$ was compared to (atomistic) Molecular Dynamics (MD) calculations 
for self affine fractal surfaces (with fractal dimension $D_{\rm f} = 2.2$).
The MD-result for $\alpha$ was found to be 
$\approx 14  \%$ larger than the theoretical prediction, which is very
similar to
the numerical results of Hyun et al\cite{summera4}. Thus, 
for the same fractal dimension $D_{\rm f} = 2.2$ as in the MD calculation, 
Hyun found
$\approx 13 \%$ larger $\alpha$
than predicted by the analytical theory (see Fig. \ref{Mark}). 
Very recently, M. Borri-Brunetto (private communication) has performed another 
discretized continuum mechanics (DCM) 
calculation (for exactly the same surface roughness profile as used in the MD calculation) using the
approach described in Ref. \cite{summera2}, and found $\alpha$ (at the highest magnification)
to be $\approx 20 \%$ {\it smaller} 
than the theoretical value. 
Taking into account the atomistic nature of the MD model (and the resulting problem with how to define
the contact area),
and the problem with the finite grid size in the DCM 
calculations, the agreement with the theory of Persson is remarkable good. Since the
Persson theory becomes exact as one approach complete contact, it is likely that the
theory is highly accurate for all squeezing forces $F_{\rm N}$.

\begin{figure}[htb]
\includegraphics[width=0.35\textwidth]{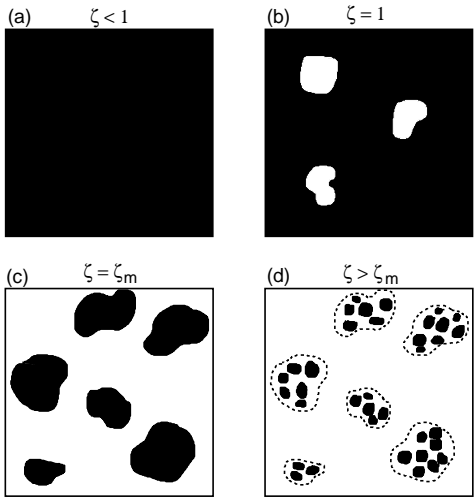}
\caption{
The contact area at increasing magnification (a)-(d). The macro-asperity
contact area (c) breaks up into smaller contact areas (d) 
as the magnification is increased.
} 
\label{newnew1}
\end{figure}

\begin{figure}[htb]
   \includegraphics[width=0.4\textwidth]{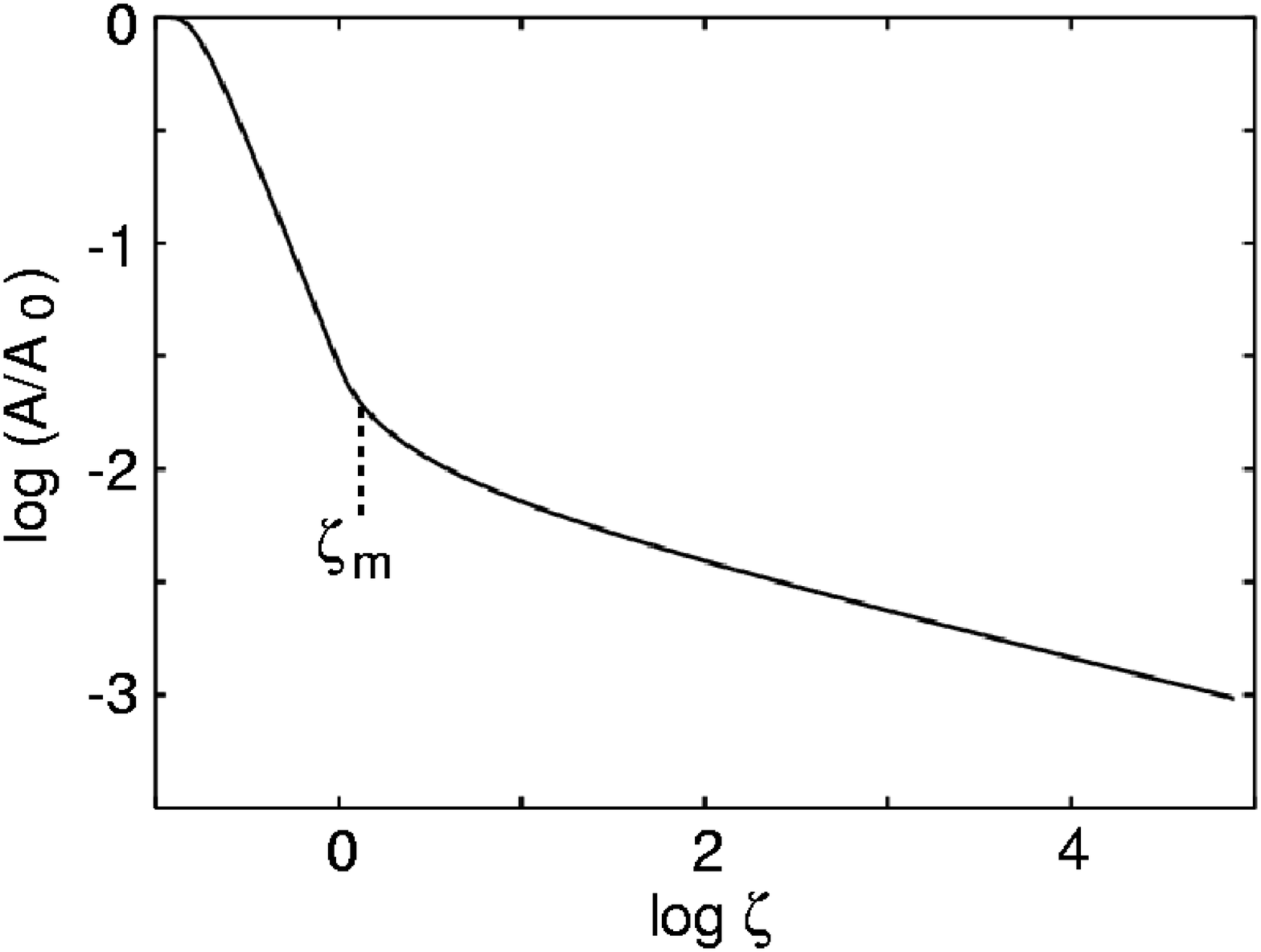}
\caption{
The logarithm of the (normalized) contact area as a function of the logarithm of
the magnification. The contact regions at the magnification
$\zeta_{\rm m}$, where the second derivative of the curve is maximal, is 
denoted as the
macroasperity contact regions. 
The elastic modulus $E=10^{11} \ {\rm Pa}$ and Poisson ratio $\nu = 0.3$. 
For the
surface ``polished steel'' with the power spectra 
shown in Fig. \ref{C.frozen} and with the (nominal) 
squeezing pressure $\sigma_0 = 10 \ {\rm MPa}$.
}
\label{1w}
\end{figure}

\vskip 0.3cm
{\bf 3.6 Macroasperity contact}

Consider the contact between two solids with nominal flat surfaces.
Assume that the surface roughness power spectrum has the qualitative form shown in
Fig. \ref{Cq1} with a roll-off wavevector $q_0$ corresponding 
to the magnification $\zeta=1$.
Fig. \ref{newnew1} shows schematically
the contact region at increasing magnification (a)-(d). 
At low magnification $\zeta < 1$ it appears as if complete contact occur between the solids.
At the magnification $\zeta \sim 1$ a few non-contact islands will be observed.
As the magnification is increased further the non-contact area will percolate giving rise
to contact islands completely surrounded by non-contact area.

Fig. \ref{1w} shows, for a typical case, 
the logarithm of the (normalized) contact area 
as a function of the logarithm of
the magnification. We denote the contact regions at the magnification
$\zeta_{\rm m}$, where the second derivative of the curve is maximal, 
as the
{\it macroasperity} contact regions. 
The macroasperity contact regions typically appear for a magnification 
somewhere in the range $\zeta_{\rm m} =q_{\rm m}/q_0 \approx
1-5$, 
where 
$A/A_0$ 
is below the (contact) percolation threshold
(which typically occur when
$A/A_0 \approx 0.3$). 
The macroasperity
contact regions [fig. \ref{newnew1}(c)] 
breaks up into smaller contact regions (d) 
as the magnification is increased.

The concept of macroasperity regions is very important in rubber
friction, in particular in the description of the influence of the flash temperature on
rubber friction\cite{Flash}.

\vskip 0.3cm
{\bf 3.7 On the validity of contact mechanics theories}

The Greenwood-Williamson theory and the theory of Bush et al (and most other
contact theories) are based on the assumption that the (average) separation between 
nearby contact regions is large enough so that the elastic deformation field 
arising from one contact region 
has a negligible influence on the
elastic deformation at all other contact regions. This assumption is strictly valid
only at low enough squeezing forces {\it and} for the magnifications $\zeta
\approx \zeta_{\rm m}$, where  
the contact regions takes the form shown schematically in Fig. \ref{newnew1}(c).
The reason for why, in general, one cannot apply these theories at higher magnification
is that for magnification $\zeta > \zeta_{\rm m}$, the macro-asperity contact regions break up into 
smaller closely spaced asperity contact regions (see Fig. \ref{newnew1}) and it is, in general, not possible to
neglect the elastic coupling between these {\it microasperity} contact regions. We note that the (average)
pressure in the macroasperity contact regions is {\it independent} of the squeezing pressure
$\sigma_0$ (but the total macroasperity contact area is proportional to
$\sigma_0$) so that the failure of the Greenwood-Williamson and Bush et al theories 
to describe the contact mechanics at high magnification
is independent of the nominal contact pressure.
The theory of Persson does (in an approximate way) account for the elastic coupling, and may hold
(approximately) for all magnifications.

\vskip 0.5 cm
{\bf 4 Elastic contact mechanics}

Consider an elastic solid with a nominal flat surfaces in contact
with a rigid, randomly rough substrate.
We assume that there is no adhesive
interaction between the solids. In this case no tensile stress 
can occur anywhere on the surfaces of the solids,
and the stress distribution
$P(\sigma, \zeta)$ must vanish for $\sigma < 0$. For this case one can show that 
$P(\sigma, \zeta)$ vanish as $\sigma \rightarrow 0$.
This can be proved in two ways. First, it is known from continuum mechanics that 
when two solids with smooth curved surfaces are in contact (without adhesion)
the stress in a contact area varies as $\xi^{1/2}$ with the distance $\xi$ 
from the boundary line of the contact area. From this follows that
$P(\sigma, \zeta) \sim \sigma$ as $\sigma \rightarrow 0$ and in particular
$P(0, \zeta)=0$, see Ref. \cite{P66}. 
For the special case of bodies with quadratic surface
profiles the stress distribution is known analytically. In particular, 
for spherical
surfaces the contact area is circular (radius $r_0$) and the pressure distribution
\cite{Jon}
$$\sigma (r) = \sigma_{\rm max} 
\left [1-\left ( {r \over r_0}\right )^2\right ]^{1/2}$$
where $\sigma_{\rm max}=3\sigma_0/2$, where $\sigma_0$ is the average 
normal stress in the contact area.
Thus, in this case
$$P(\sigma, \zeta) = {1\over \pi r_0^2} \int d^2x \ \delta (\sigma - \sigma(r))= 
{2\sigma \over \sigma^2_{\rm max}}\eqno(23)$$
which vanishes linearly with $\sigma$.

A second proof follows from (17), and the observation that at any magnification the
contact pressure integrated over the (apparent) area of contact must be equal to the load. 
Thus, multiplying (17) with $\sigma $ and integrating over $\sigma$ gives
$${\partial \over \partial \zeta} \int_0^\infty d\sigma \ \sigma P(\sigma, \zeta) =
\int_0^\infty d\sigma \ f(\zeta) \sigma {\partial^2 P \over \partial \sigma^2}$$ 
$$= - f(\zeta) \int_0^\infty d\sigma \ {\partial P \over \partial \sigma} (\sigma ,\zeta)
= f(\zeta) P(0,\zeta)\eqno(24)$$
where we have performed a partial integration and assumed that
$\sigma \partial P(\sigma, \zeta)/\partial \sigma \rightarrow 0$ as $\sigma \rightarrow \infty$
(which holds because $P(\sigma, \zeta)$ decreases exponentially or faster with increasing $\sigma$, 
for large $\sigma$).
Since
$$\int_0^\infty d\sigma \ \sigma P(\sigma, \zeta) = \sigma_0$$
for all magnifications we get
$${\partial \over \partial \zeta} \int_0^\infty d\sigma \ \sigma P(\sigma, \zeta) =0\eqno(25)$$
Thus, since $f(\zeta) \neq 0$ we get from (24) and (25) that $P(0,\zeta)=0$. 
Hence, (24) {\it together with the observation that the load must be carried by the
area of (apparent) contact at any magnification, imply} $P(0,\zeta)=0$.
The fact that this (exact) result follows from (24) 
is very satisfactory and is not trivial
since equation (24) is not exact
when partial contact occur at the interface. Thus, the fact that (24) is consistent with the
correct boundary condition lend support for the usefulness 
of the present contact mechanics theory.

Integrating (24) over $\sigma$ gives
$${\partial \over \partial \zeta} \int_0^\infty d\sigma \ P(\sigma, \zeta) =
\int_0^\infty d\sigma \ f(\zeta) {\partial^2 P \over \partial \sigma^2} 
= - f(\zeta) {\partial P \over \partial \sigma} (0,\zeta)$$
and since 
$$\int_0^\infty d\sigma \ P(\sigma, \zeta) =
{A(\zeta) \over A_0 } \equiv P(\sigma)$$ 
we get
$$A' = 
- A_0 f(\zeta) {\partial P \over \partial \sigma} (0,\zeta)\equiv A'_{\rm el}$$
where the index ``el'' indicate that the solids deform elastically in the area of contact.
Since the total area (projected on the $xy$-plane) is conserved
$A_{\rm el}+A_{\rm non} = A_0$, where $A_{\rm non}$ denote the non-contact area, 
we get $A'_{\rm el}=-A'_{\rm non}$ so that 
$$A'_{\rm non}= A_0 f(\zeta) {\partial P \over \partial \sigma} (0,\zeta)\eqno(26)$$
In analogy with heat current, we can interpret this equation as describing the {\it flow}
of area from contact into non-contact with increasing magnification.

\begin{figure}[htb]
   \includegraphics[width=0.4\textwidth]{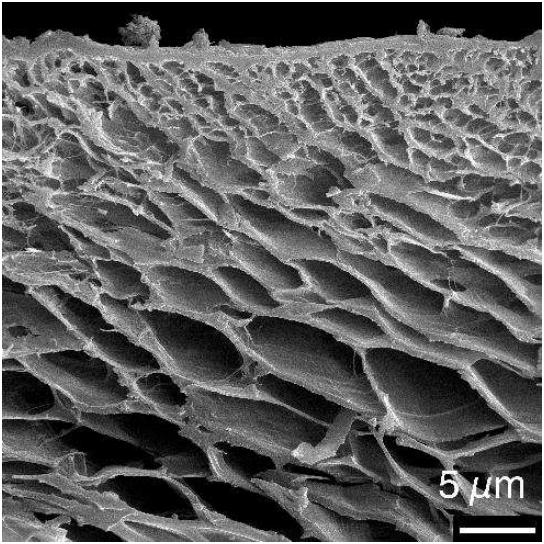}
\caption{
Vertical cut through the adhesion pad of cicada. The foam-like structure of the adhesion pad makes it
elastically soft, which is necessary for the pad to adhere to rough surfaces. In addition, the
elastic properties are graded\cite{foam} 
which may also be important for the adhesion (see Sec. 8.5). 
Adapted from Ref. \cite{Sherg}.
}
\label{Gorb}
\end{figure}

\vskip 0.5 cm
{\bf 5 Elastic contact mechanics with graded elasticity}

In many applications the elastic properties of solids are not constant in space but varies with
the distance $z$ from the surface into the solid. Thus, for example, many solids have modified surface
layers, e.g., thin 
oxide layers with different elastic properties than the bulk.
In these cases the elastic modulus $E=E(z)$ will depend on the distance $z$ from the surface.
The same is true for many biological systems where natural selection has optimized the surface
properties for maximum functionality. Thus, for example, the adhesion pads (see Fig. \ref{Gorb})
on many
insects consist of graded materials which may exhibit robust 
flaw tolerant adhesion (see Sec. 8.5).
It is therefore of great interest to study contact mechanics, both with and without adhesion, 
for solids with elastic properties
which depend on the distance from the surface into the solid.  

The theory developed in Ref.\cite{P8,P1}
can be (approximately) applied also to 
study contact mechanics for solids with graded elastic properties. Thus, the deformation field
in a solid in a response to a surface stress distribution which varies parallel to the surface 
with the wavelength $\lambda$ (wave vector $q=2\pi /\lambda$) 
extend into the solid a distance of order $\sim 1/q$, and is 
approximately of the form $\sim {\rm exp} (-qz)$. 
Thus, if we replace $E$ in (18) with $\tilde E(\zeta)$ 
defined by
$$\tilde E(\zeta) = {\int_0^\infty dz \ E(z) {\rm exp}(-qz) \over
\int_0^\infty dz \ {\rm exp}(-qz)}=
q\int_0^\infty dz \ E(z) {\rm exp}(-qz)$$
where $q=q_0\zeta$, we account for the fact that the 
effective elastic modulus depends on the wavelength of the roughness profile. 
This approach should work best if the
spatial variation in $E(z)$ is weak, e.g., $E'(z)/E(z) < q$ for $z\sim 1/q$. 
The same approach can be used when adhesion is included, but then the
elastic modulus which enter in the detachment stress, and in the expression for the elastic energy,
must also be replaced by the expression $\tilde E(\zeta)$ defined above.
The limiting case of a thin elastic plate can be treated using the formalism developed in Ref.
\cite{LizardP,LizardCP}.

\vskip 0.5 cm
{\bf 6 Elastoplastic contact mechanics: constant hardness}

We assume ideal elastoplastic solids.
That is, the solids deform elastically until the normal stress
reach the yield stress $\sigma_{\rm Y}$ (or rather the indentation hardness, which is $\sim 3$ times
the yield stress of the material in simple compression),
at which point plastic flow occur. The stress in the plastically deformed contact area is assumed 
to be $\sigma_{\rm Y}$ (e.g., no strain hardening). Contact mechanics for elastoplastic solids
was studied in Ref. \cite{P8}, where the stress distribution was calculated from (17) using the
boundary conditions $P(0,\zeta)=0$ and $P(\sigma_{\rm Y}, \zeta)=0$. The first 
(elastic detachment) boundary condition was 
proved above and here we show that the second (plastic yield) boundary condition
follows from Eq. (17) together with 
the condition that at any magnification the
contact pressure integrated over the (apparent) area of contact must be equal to the load. 
In this case the load is carried in part by the plastically deformed contact area.

Integrating (17) from $\sigma=0$ to $\sigma_{\rm Y}$ gives
$${\partial \over \partial \zeta} \int_0^{\sigma_{\rm Y}} d\sigma \ P(\sigma, \zeta) =
\int_0^{\sigma_{\rm Y}} d\sigma \ f(\zeta) {\partial^2 P \over \partial \sigma^2}$$ 
$$= - f(\zeta) {\partial P \over \partial \sigma} (0,\zeta)
+f(\zeta) {\partial P \over \partial \sigma} (\sigma_{\rm Y},\zeta)
\eqno(27)$$
Since
$$\int_0^{\sigma_{\rm Y}} d\sigma \ P(\sigma, \zeta) = P(\zeta) 
= {A_{\rm el}(\zeta) \over A_0}$$
where $A_{\rm el}$ is the ``elastic contact area'',
we can write (27) as
$$A'_{\rm el} (\zeta) =
- A_0f(\zeta) {\partial P \over \partial \sigma} (0,\zeta)+
A_0f(\zeta) {\partial P \over \partial \sigma} (\sigma_{\rm Y},\zeta)$$
$$=-A'_{\rm non}(\zeta)-A'_{\rm pl}(\zeta)\eqno(28)$$
where we have used (26), and where
$$A'_{\rm pl}(\zeta)=
-A_0f(\zeta) {\partial P \over \partial \sigma} (\sigma_{\rm Y},\zeta)\eqno(29)$$
is the contact area where plastic deformation has occurred.
Thus, (28) is just a statement about conservation of area (projected on the $xy$-plane):
$$A'_{\rm el}(\zeta)+A'_{\rm pl}(\zeta )+A'_{\rm non}(\zeta ) = 0,$$
or after integration
$$A_{\rm el}+A_{\rm pl}+A_{\rm non} = A_0$$

Next, let us multiply (17) with $\sigma$ and integrate over $\sigma$:
$${\partial \over \partial \zeta} \int_0^{\sigma_{\rm Y}} d\sigma \ \sigma P(\sigma, \zeta) =
\int_0^{\sigma_{\rm Y}} d\sigma \ f(\zeta) \sigma {\partial^2 P \over \partial \sigma^2}$$ 
$$= f(\zeta) \sigma_{\rm Y}{\partial P \over \partial \sigma}(\sigma_{\rm Y},\zeta)- 
f(\zeta) \int_0^{\sigma_{\rm Y}} d\sigma \ {\partial P \over \partial \sigma} (\sigma,\zeta)$$
$$= f(\zeta) \left [ \sigma_{\rm Y}{\partial P \over \partial \sigma}(\sigma_{\rm Y},\zeta)- 
P(\sigma_{\rm Y},\zeta)\right ]\eqno(30)$$
Since
$$F_{\rm el}(\zeta) = A_0
\int_0^{\sigma_{\rm Y}} d\sigma \ \sigma P(\sigma, \zeta)$$
is the normal force carried by the elastically deformed contact area, (30) gives 
$$F'_{\rm el}(\zeta) = 
A_0 f(\zeta) \sigma_{\rm Y}{\partial P \over \partial \sigma}(\sigma_{\rm Y},\zeta)- 
A_0f(\zeta)P(\sigma_{\rm Y},\zeta)$$
Using (29) this gives
$$F'_{\rm el}(\zeta) = 
-\sigma_{\rm Y}A'_{\rm pl}- 
A_0f(\zeta)P(\sigma_{\rm Y},\zeta)\eqno(31)$$
Since the normal stress in the contact region where plastic yield has occurred is equal to
$\sigma_{\rm Y}$ it follows that the plastically yielded contact regions will carry the load
$F_{\rm pl}=\sigma_{\rm Y}A_{\rm pl}$. Thus, since $\sigma_{\rm Y}$ is assumed independent of
$\zeta$ we get 
$F'_{\rm pl}=\sigma_{\rm Y}A'_{\rm pl}$ and
(31) can be written as
$$F'_{\rm el}(\zeta) = 
-F'_{\rm pl}(\zeta)- 
A_0f(\zeta)P(\sigma_{\rm Y},\zeta)\eqno(32)$$
But the total load $F_{\rm N}=\sigma_0A_0= F_{\rm el}(\zeta)+F_{\rm pl}(\zeta)$ 
must be independent of the magnification
so that 
$F'_{\rm el}(\zeta)+F'_{\rm pl}(\zeta)=0$. Comparing this with (32) gives
the boundary condition $P(\sigma_{\rm Y},\zeta)=0$.

In Ref. \cite{P8} the ``diffusion equation'' (17) was solved 
analytically with the boundary conditions
$P(0,\zeta)=0$ and $P(\sigma_{\rm Y},\zeta)=0$. 
Qualitative we may state that with increasing magnification the contact area diffuses over the
$\sigma=0$ boundary into no-contact, and over the $\sigma=\sigma_{\rm Y}$  boundary into plastic
contact.  

\begin{figure}[htb]
   \includegraphics[width=0.4\textwidth]{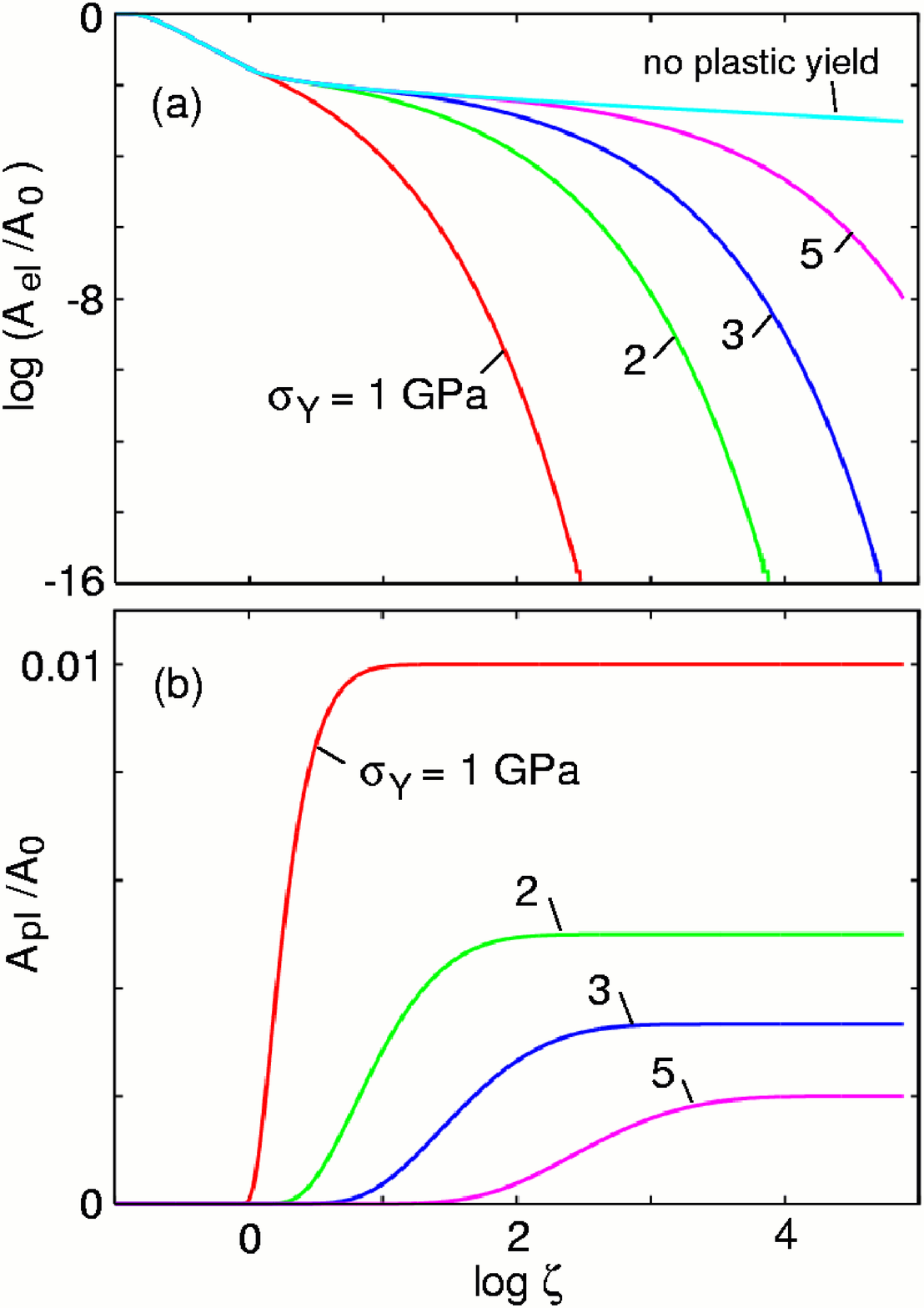}
\caption{
The dependence of (a) the logarithm of the (normalized) elastic contact area 
and (b) the (normalized) plastic contact area
on the logarithm of the magnification.
The elastic modulus $E=10^{11} \ {\rm Pa}$, Poisson ratio $\nu = 0.3$ and for several
yield stresses indicated in the figure.
For the
surface ``polished steel'' with the power spectra 
shown in Fig. \ref{C.frozen} and with the (nominal) 
squeezing pressure $\sigma_0 = 10 \ {\rm MPa}$.
}
\label{1z}
\end{figure}

As an illustration Fig. \ref{1z} shows how 
(a) the logarithm of the (normalized) elastic contact area, 
and (b) the (normalized) plastic contact area depend
on the logarithm of the magnification.
The calculation is for the
surface ``polished steel'' with the power spectrum 
shown in Fig. \ref{C.frozen} and with the (nominal) 
squeezing pressure $\sigma_0 = 10 \ {\rm MPa}$.
The elastic modulus $E=10^{11} \ {\rm Pa}$, Poisson ratio $\nu = 0.3$ and for several
yield stresses indicated in the figure.
Note that with increasing magnification, 
the plastically yielded contact area $A_{\rm pl}/A_0$ 
increases continuously towards the limiting value
$\sigma_0/\sigma_{\rm Y}$. 

\vskip 0.5 cm
{\bf 7 Elastoplastic contact mechanics: size-dependent hardness}

Many materials exhibit a penetration hardness which depend on
the size of the indentation. That is, $\sigma_{\rm Y}$ is a function of $\zeta$.
In this case
$${\partial \over \partial \zeta} \int_0^{\sigma_{\rm Y}} d\sigma \ P(\sigma, \zeta) =
\sigma'_{\rm Y}(\zeta) P(\sigma_{\rm Y},\zeta)+
\int_0^{\sigma_{\rm Y}} d\sigma \ f(\zeta) {\partial^2 P \over \partial \sigma^2}$$ 
$$= \sigma'_{\rm Y}(\zeta)P(\sigma_{\rm Y},\zeta)- f(\zeta) {\partial P \over \partial \sigma} (0,\zeta)
+f(\zeta) {\partial P \over \partial \sigma} (\sigma_{\rm Y},\zeta)
$$
which differ from (27) by the term $\sigma'_{\rm Y}(\zeta)P(\sigma_{\rm Y},\zeta)$. 
In this case area conservation require the definition
$$A'_{\rm pl}(\zeta)=
-A_0\sigma'_{\rm Y}(\zeta) P(\sigma_{\rm Y},\zeta)-A_0f(\zeta) {\partial P \over \partial \sigma} 
(\sigma_{\rm Y},\zeta)\eqno(33)$$
A similar ``extra'' term
will appear in (30) but (31) will remain unchanged.
Thus, since
$F'_{\rm pl}(\zeta)= \sigma'_{\rm Y}(\zeta)A_{\rm pl}(\zeta)  
+\sigma_{\rm Y}(\zeta)A'_{\rm pl}(\zeta)$, (31) takes the form  
$$F'_{\rm el}(\zeta) = 
-F'_{\rm pl}(\zeta)+ 
\sigma'_{\rm Y}(\zeta)A_{\rm pl}(\zeta)-  
A_0f(\zeta)P(\sigma_{\rm Y},\zeta)$$
Hence, in this case
$$\sigma'_{\rm Y}(\zeta)A_{\rm pl}(\zeta)-  
A_0f(\zeta)P(\sigma_{\rm Y},\zeta)=0\eqno(34)$$
This equation together (33) gives the correct boundary condition when $\sigma_{\rm Y}(\zeta)$
depend on $\zeta$.

\vskip 0.5 cm
{\bf 8 Elastic contact mechanics with adhesion}

In this section I first derive and discuss the boundary conditions 
for the stress distribution relevant for adhesive contact between solids.
I also derive an improved expression for the elastic energy stored at the interface
in an adhesive contact. Numerical results are presented to illustrate the theory.
Finally, I present a few comments related to flaw tolerant adhesion, adhesion involving
viscoelastic solids, and adhesion in biological systems. 

\vskip 0.5 cm
{\bf 8.1 Boundary condition and physical interpretation}

When adhesion is included,
the stress distribution function $P(\sigma , \zeta)$ must vanish for $\sigma < - \sigma_{\rm a}$,
where $\sigma_{\rm a} (\zeta)$ is the highest tensile stress possible at the interface when the system is
studied at the magnification $\zeta$. 
We will now prove the boundary condition 
$P(-\sigma_{\rm a}(\zeta),\zeta)=0$. 
When the adhesion is neglected $\sigma_{\rm a} = 0$ and 
$P(0,\zeta)=0$ which is the limiting case studied in Sec. 4. 

In the present case, the (normalized) area of (apparent) contact 
when the system is studied at the resolution $\zeta$ is given by
$$P(\zeta)={A(\zeta)\over \ A_0} = \int_{-\sigma_{\rm a}(\zeta)}^\infty d \sigma 
\ P(\sigma, \zeta)\eqno(35)$$
where $A_0$ is the nominal contact area.
The area of real contact is obtained from (35) at the highest (atomic) resolution,
corresponding to the magnification $\zeta_1=L/a$,
where $a$ is an atomic distance.

\begin{figure}[htb]
   \includegraphics[width=0.3\textwidth]{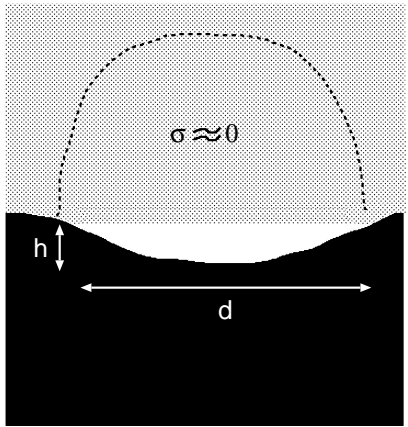}
\caption{
An elastic solid (dotted area) in adhesive contact with a hard
solid (black area) with surface roughness. The elastic solid will deform
and fill out the substrate ``cavity'' only if the gain in adhesive energy $\sim \gamma d^2$ 
is of order (or larger than) the elastic
energy stored in the deformation field. The deformation field extend a distance
$d$ into the solid, and the strain is of order $h/d$ giving the elastic energy of order
$d^3 E (h/d)^2$. The detachment stress $\sigma_{\rm a} = E h/d$ is defined as the
stress when $U_{\rm ad} = U_{\rm el}$ or $\gamma d \approx Eh^2$ so that
$\sigma_{\rm ad} \approx (\gamma E/d)^{1/2}$. Since $d$ is of order the wavelength of
the surface roughness we get (36) to within a factor of order unity.
}
\label{1u}
\end{figure}

The detachment stress $\sigma_{\rm a}$ can be determined using the theory of cracks
which gives (see Fig. \ref{1u}) 
$$\sigma_{\rm a} (\zeta) \approx \left ({\alpha \gamma_{\rm eff} (\zeta) E 
q \over 1-\nu^2}\right )^{1/2}\eqno(36)$$ 
where $\alpha$ is a number of order unity (we use $\alpha = 9/8$) and
$\gamma_{\rm eff} (\zeta)$ is the (effective) interfacial binding energy when the system is studied
at the magnification $\zeta$. At the highest magnification (corresponding to atomic 
resolution) $\gamma_{\rm eff}= \Delta \gamma$, where $\Delta \gamma = \gamma_1+\gamma_2-\gamma_{12}$ 
is the change in the energy (per unit area) when two {\it flat} 
surfaces of the solids are brought into adhesive contact. 
The effective interfacial energy at the magnification $\zeta$ is determined by the equation
$$A^*(\zeta) \gamma_{\rm eff}(\zeta ) =A^*(\zeta_1)\Delta \gamma - U_{\rm el}(\zeta)\eqno(37)$$ 
where $A^*(\zeta)$ denote the total contact area
which may differ from the projected contact area $A(\zeta)$.
$U_{\rm el}(\zeta)$ is the elastic energy stored at the interface due to the elastic 
deformation of the solid on length scales shorter than $\lambda = L/\zeta$, necessary in order to bring 
the solids into adhesive contact. 

\begin{figure}[htb]
   \includegraphics[width=0.4\textwidth]{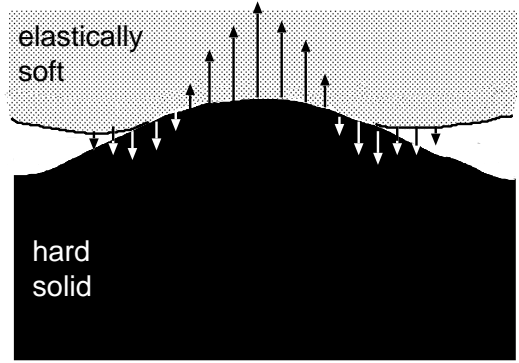}
\caption{
The stress distribution in an asperity contact region. The stress is tensile close to the
boundary line of the contact area. An attractive wall-wall interaction occur also outside the
area of real contact and this interaction determines the detachment stress $\sigma_{\rm a}$
and the work of adhesion
$\gamma$.
}
\label{Pres}
\end{figure}

Using (17) we get
$${\partial \over \partial \zeta} \int^\infty_{-\sigma_{\rm a}} d\sigma \ P(\sigma, \zeta) =
\sigma'_{\rm a}(\zeta) P(-\sigma_{\rm a},\zeta)+
\int^\infty_{-\sigma_{\rm a}} d\sigma \ f(\zeta) {\partial^2 P \over \partial \sigma^2}$$ 
$$= \sigma'_{\rm a}(\zeta)P(-\sigma_{\rm a},\zeta)
-f(\zeta) {\partial P \over \partial \sigma} (-\sigma_{\rm a},\zeta)
$$
or
$$A'(\zeta) = A_0\sigma'_{\rm a}(\zeta)P(-\sigma_{\rm a},\zeta)
-A_0f(\zeta) {\partial P \over \partial \sigma} (-\sigma_{\rm a},\zeta)
\eqno(38)$$
Next, using (17) we get
$${\partial \over \partial \zeta} \int^\infty_{-\sigma_{\rm a}} d\sigma \ \sigma P(\sigma, \zeta) =
-\sigma'_{\rm a}(\zeta) \sigma_{\rm a}(\zeta) P(-\sigma_{\rm a},\zeta)$$
$$+ 
\int^\infty_{-\sigma_{\rm a}} d\sigma \ f(\zeta) \sigma {\partial^2 P \over \partial \sigma^2}$$ 
$$= 
-\sigma'_{\rm a}(\zeta) \sigma_{\rm a}(\zeta) P(-\sigma_{\rm a},\zeta)$$
$$+ 
f(\zeta) \sigma_{\rm a}{\partial P \over \partial \sigma}(-\sigma_{\rm a},\zeta)- 
f(\zeta) \int^\infty_{-\sigma_{\rm a}} d\sigma \ {\partial P \over \partial \sigma} (0,\zeta)$$
$$= 
-\sigma'_{\rm a}(\zeta) \sigma_{\rm a}(\zeta) P(-\sigma_{\rm a},\zeta)$$
$$+ 
f(\zeta) \left [ \sigma_{\rm a}{\partial P \over \partial \sigma}(-\sigma_{\rm a},\zeta)+
P(-\sigma_{\rm a},\zeta)\right ]$$
Thus, the normal force acting in the area of contact
satisfies
$$F'(\zeta)=
-A_0\sigma'_{\rm a}(\zeta) \sigma_{\rm a}(\zeta) P(-\sigma_{\rm a},\zeta)$$
$$+ 
A_0f(\zeta) \left [ \sigma_{\rm a}{\partial P \over \partial \sigma}(-\sigma_{\rm a},\zeta)+
A_0P(-\sigma_{\rm a},\zeta)\right ]$$
Using (38) this gives
$$F'(\zeta)=
-\sigma_{\rm a}(\zeta) A'(\zeta)
+A_0P(-\sigma_{\rm a},\zeta)\eqno(39)$$
Now, note that since $A'(\zeta) < 0$, {\it both} terms on the right hand side of (39) are positive.
This implies that the {\it normal force} $F(\zeta)$ {\it acting in the area of contact cannot, in
general, be
equal to the external load} $F_{\rm N}=A_0\sigma_0$. This remarkable result is possible only if there
is an attractive interaction between the non-contacting surfaces. However, this is exactly what one expect
in the present case for the following reason: The (effective) surface energy $\gamma_{\rm eff}$
is determined by the work to separate the surfaces. Since the stress field in our 
model is everywhere finite, in particular the detachment stress $\sigma_{\rm a}$ is finite, the 
surface forces must extend over a finite extent in order for the work of separation to remain non-zero.
Only when the stresses are allowed to become infinite is it possible to have a finite work of adhesion
with contact interaction (i.e., interaction of infinitesimal extent) $\gamma = 0\times \infty$.
{\it Since the detachment stress is finite in our model, the attractive 
interaction between the walls must have a
finite extent and exist also outside the area of contact}, see Fig. \ref{Pres}.
Thus, (39) cannot be used to determine the boundary condition at $\sigma=-\sigma_{\rm a}$,
but describes how the force acting in
the contact area increases with increasing magnification, as a result of the adhesional ``load''. 
However, the boundary condition 
can be determined using the first method described in Sec. 4, by considering
the stress distribution close to the boundary line between the contact 
and non-contact area, which we refer to as the detachment line.
The detachment line can be considered as a crack edge, and from the general theory of cracks one
expect the tensile stress to diverge as $\xi^{-1/2}$ with the distance $\xi$ from the detachment line.
However, this is only true when the interaction potential between the surfaces is infinitesimally short
ranged, which is not the case in the present study (see above),
where the stress close to the boundary line (in the crack tip process zone)
will be of the form $\sigma = \sigma_{\rm a} +
C\xi^{1/2}$, where $\xi$ is the distance from the boundary line. 
This gives $d\sigma/d\xi \sim \xi^{-1/2}$ and following the argument presented
in Sec. 4 (and in Ref. \cite{P66}) this imply that the stress distribution will vanish
linearly with $\sigma+\sigma_{\rm a}$ as $\sigma \rightarrow -\sigma_{\rm a}$ 
so that $P(-\sigma_{\rm a},\zeta) = 0$. For the special case of a circular
asperity contact region formed between two spherical surfaces
in adhesive contact,
Greenwood and Johnson\cite{GJ} 
have presented an elegant approach, which I use in Appendix B to calculate the
stress distribution in the area of real contact. 
I find that the stress distribution is Hertzian-like but with 
a tail of the form $\sim (\sigma+\sigma_{\rm a})$ extending to negative (tensile) stress. 

Note that with $P(-\sigma_{\rm a},\zeta)=0$, Eq. (39) takes the form
$$F'(\zeta)=
-\sigma_{\rm a}(\zeta) A'(\zeta)\eqno(40)
$$
Let us consider the simplest case where $\sigma_{\rm a}$ is independent of the magnification $\zeta$.
In this case (40) gives
$${d\over d\zeta} \left [ F(\zeta)+\sigma_{\rm a}A(\zeta)\right ] = 0$$
or
$$F(\zeta)+\sigma_{\rm a}A(\zeta)= (\sigma_0+\sigma_{\rm a}) A_0\eqno(41)$$
where we have used that $F(1)=\sigma_0A_0$ and $A(1)=A_0$. Thus,
if we assume that $A(\zeta)<<A_0$, as is typically the case for elastically hard solids
and large magnification, then
$$F(\zeta) \approx (\sigma_0+\sigma_{\rm a}) A_0\eqno(42)$$
i.e., the normal force in the contact area is the sum of the applied load $\sigma_0A_0$ and the adhesion
load $\sigma_{\rm a}A_0$. This result was derived in Ref. \cite{P66} 
using a different approach, and has often been
used\cite{Derja}. Note, however,
that in most practical situations 
$\sigma_{\rm a}$ will depend strongly on the magnification and the relation (42) 
is then no longer valid. One exception may be systems with a suitable chosen graded elasticity,
see
Sec. 8.5.

\vskip 0.3cm
{\bf 8.2 Elastic energy}

When complete contact occur between the solids on all
length scales, $U_{\rm el}=U_{\rm el}^0$ with\cite{P1} 
$$U_{\rm el}^0= {A_0 E\over 4(1-\nu^2)}\int d^2q \ qC(q)$$
$$ = 
{A_0 E\over 4(1-\nu^2)} 2\pi q_0^3 \int_1^{\zeta_1} d\zeta \ \zeta^2 C(q_0\zeta)$$
where $q_1 = \zeta_1 q_0$ is the largest surface roughness wave vector. In Ref. \cite{P1}
we accounted for the fact that only partial contact occur at the interface by using 
$$U_{\rm el}= 
{A_0 E\over 4(1-\nu^2)} 2\pi q_0^3 \int_1^{\zeta_1} d\zeta \ \zeta^2 C(q_0\zeta)P(\zeta)\eqno(43)$$
where $P(\zeta)=A(\zeta)/A_0$ is the ratio between the area of real contact (projected
on the $xy$-plane), $A(\zeta)$, when the system is studied at the magnification $\zeta$, and the
nominal contact area $A_0$. 
Here we will use a (slightly) 
more accurate expression for $\gamma_{\rm eff}$ determined as follows.

First note that
$$P(\zeta)={A(\zeta)\over A_0}={\int_{-\sigma_c(\zeta)}^\infty d\sigma \ P(\sigma, \zeta) \over
\int_{-\infty}^\infty d\sigma \ P_0(\sigma, \zeta)}$$
where $P_0(\sigma,\zeta)=P_0(\sigma)$ is the stress distribution assuming 
complete contact at all magnifications.

The factor $P(\zeta)$ in Eq. (43) arises from the assumption that the elastic energy which 
would be stored (per unit area) in the region
which we denote as ``detached area'', if this area would not detach,
is the same as the average elastic energy per unit area assuming complete contact. 
However, in reality the stress and hence the elastic energy which would be stored 
in the vicinity of the ``detached area'', if no detachment would occur, is higher than the average
(that's why the area detach). Since the stored elastic energy is a quadratic
function of the surface stress, we can take into account this ``enhancement'' 
effect by replacing $P(\zeta)$ in (43) with
$$\tilde P(\zeta)={\int_{-\sigma_c(\zeta)}^\infty d\sigma \ (\sigma - \sigma_0)^2 P(\sigma, \zeta) \over
\int_{-\infty}^\infty d\sigma \ (\sigma-\sigma_0)^2 P_0(\sigma, \zeta)}$$
or
$$\tilde P(\zeta) =     
{\langle (\sigma-\sigma_0)^2\rangle_\zeta \over
\langle (\sigma-\sigma_0)^2\rangle_\zeta^0}\eqno(44)$$
where in the denominator $\langle ... \rangle$ is calculated assuming complete contact.
Note that $\tilde P(\zeta)$ takes into account {\it both} the decrease in the contact area
with increasing magnification, and the fact that the detached regions are the high tensile stress
regions where $-\sigma > \sigma_c(\zeta)$. Thus, $\tilde P(\zeta) < P(\zeta)$ and the amount of stored elastic
energy at the interface will be smaller when using the $\tilde P(\zeta )$-function 
instead of the $P(\zeta )$-function. 
The $\tilde P(\zeta)$ function can be written in a convenient way, involving only $P(\zeta)$ 
rather than the distribution function $P(\sigma,\zeta)$, as follows. 
Let us write
$$\langle (\sigma-\sigma_0)^2\rangle_\zeta = F(\zeta)-2\sigma_0 G(\zeta) 
+\sigma_0^2 P(\zeta)\eqno(45)$$ 
Consider first the integral
$$F(\zeta ) = \langle \sigma^2 \rangle_\zeta = 
\int_{-\sigma_c(\zeta )}^\infty d\sigma \ \sigma^2 P(\sigma, \zeta)$$
Using (17) and that $P(\sigma_c(\zeta),\zeta)=0$ we get
$$F'(\zeta ) = 
\int_{-\sigma_c(\zeta )}^\infty d\sigma \ \sigma^2 {\partial P\over \partial \zeta}$$
$$=f(\zeta) \int_{-\sigma_c(\zeta )}^\infty d\sigma \ \sigma^2 {\partial^2 P\over \partial \sigma^2}$$
Performing two partial integrations and using (35) gives
$$F'(\zeta)=f(\zeta) \left [ -\sigma_c^2(\zeta ){\partial P\over \partial \sigma}(-\sigma_c(\zeta),\zeta)
+2P(\zeta)\right ]\eqno(46)$$
Integrating (17) over $\sigma$ gives
$$P'(\zeta) = -f(\zeta) {\partial P \over \partial \sigma} (-\sigma_c(\zeta),\zeta)$$
Substituting this in (46) gives  
$$F'(\zeta) =\sigma_c^2(\zeta ) P'(\zeta)
+2f(\zeta) P(\zeta)$$
In a similar way one can show that
$$G'(\zeta ) = {d\over d\zeta} \langle \sigma \rangle_\zeta = 
-\sigma_c P'(\zeta)\eqno(47)$$
Thus
$${d\over d\zeta} \langle (\sigma-\sigma_0)^2\rangle_\zeta = 
F'(\zeta)-2\sigma_0 G'(\zeta) +\sigma_0^2 P'(\zeta)$$
$$= 
(\sigma_c(\zeta )+\sigma_0)^2 P'(\zeta)
+2f(\zeta) P(\zeta)$$
Integrating this equation 
gives 
$$\langle (\sigma-\sigma_0)^2 \rangle_\zeta  
=\int_1^\zeta d\zeta' \left [ (\sigma_c(\zeta')+\sigma_0)^2 P'(\zeta')
+2f(\zeta') P(\zeta')\right ]\eqno(48)$$
Now, assume that complete contact occur at the interface at all magnifications.
In this case $P(\zeta)\equiv 1$ and (48) takes the form
$$\langle (\sigma-\sigma_0)^2 \rangle_\zeta^0  =
\int_1^\zeta d\zeta' 2f(\zeta')\eqno(49)$$
Thus, 
using (48) and (49) gives
$$
\tilde P(\zeta) 
=
{\int_1^\zeta d\zeta' \sigma_c^2(\zeta' ) P'(\zeta') \over
2\int_1^\zeta d\zeta' f(\zeta')}+ 
{\int_1^\zeta d\zeta' \ f(\zeta') P(\zeta')
\over
\int_1^\zeta d\zeta' f(\zeta')} \eqno(50)$$

\begin{figure}[htb]
   \includegraphics[width=0.45\textwidth]{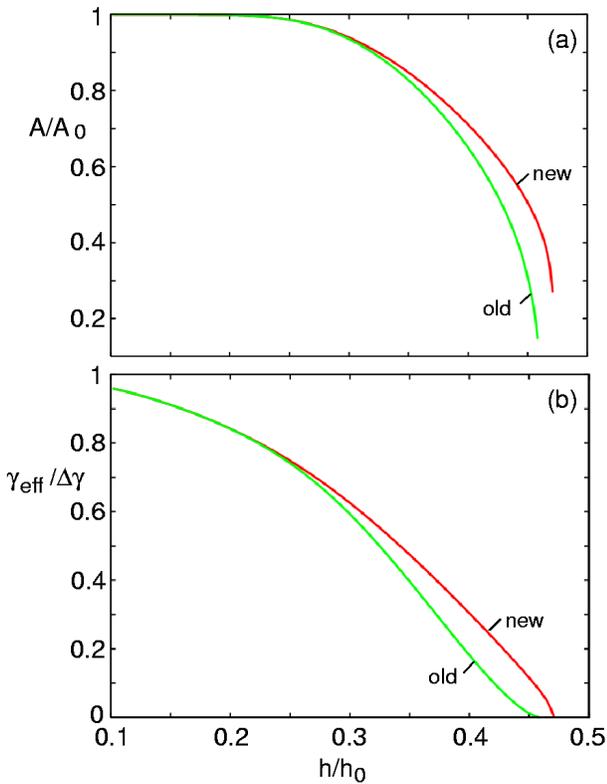}
\caption{
(a) The relative contact area 
at the highest magnification $A(\zeta_1)/A_0$, and (b) the (normalized) 
macroscopic effective interfacial energy per unit area
$\gamma_{\rm eff}(1)/\Delta \gamma$ as a function of $h/h_0$. 
We show the results using both the ``old'' form of the elastic energy and the improved
form where $P(\zeta)$ is replaced by $\tilde P(\zeta)$. 
For an
elastic solid with a flat surface in adhesive contact 
(at vanishing applied pressure $\sigma_0 = 0$)
with a hard, randomly rough
self affine fractal surface with the root-mean-square roughness $h_0=0.5 \ {\rm \mu m}$
and the fractal dimension $D_{\rm f} = 2.2$. 
The long and short distance cut-off wavevectors $q_0=10^5 \ {\rm m^{-1}}$
and $q_1=10^{10} \ {\rm m^{-1}}$, respectively.
The elastic solid has the Young's modulus $E=10 \ {\rm MPa}$
and the Poisson ratio $0.5$. The interfacial energy per unit area $\Delta \gamma =
0.05 \ {\rm J/m^2}$. 
}
\label{A.gamm.hh0}
\end{figure}

\begin{figure}[htb]
   \includegraphics[width=0.45\textwidth]{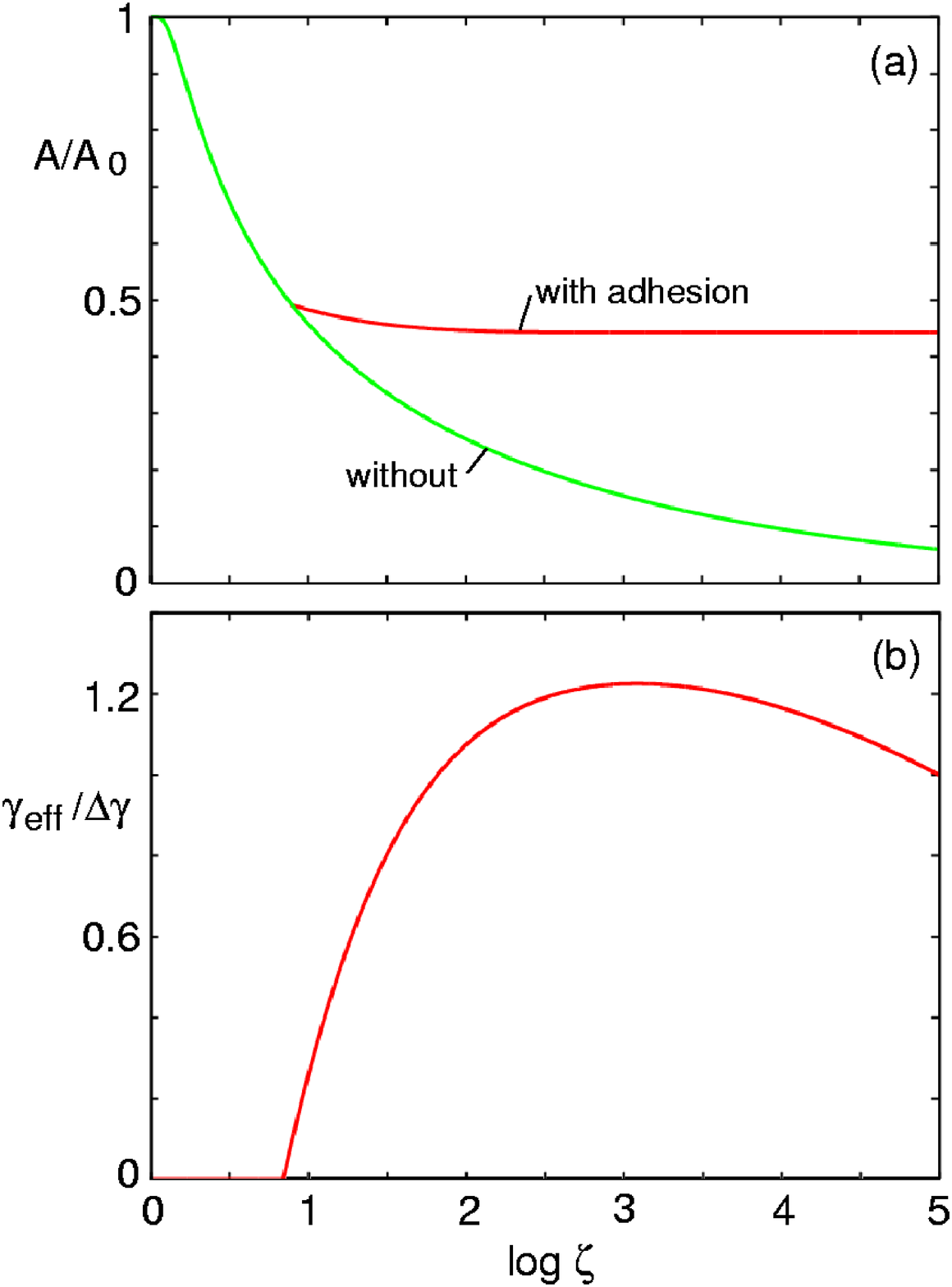}
\caption{
(a) The relative contact area $A(\zeta)/A_0$ and (b) the (normalized) 
effective interfacial energy per unit area
$\gamma_{\rm eff}(\zeta) /\Delta \gamma$ as a function of magnification. 
For the nominal pressure $\sigma_0 = 0.5 \ {\rm MPa}$. For the same system as in
Fig. \ref{A.gamm.hh0}.
}
\label{A.zeta}
\end{figure}

\vskip 0.5 cm
{\bf 8.3 Numerical results}

In many applications surface roughness occur over many decades in length scale,
e.g., from cm to nm (7 decades in length scale). In this case it is not possible to
perform (numerically) integrals over the magnification directly, 
but the integrals are easily performed 
if one change the integration variable $\zeta = {\rm exp}(\mu)$ so that  
$$\int_1^{\zeta_1} d\zeta \ F(\zeta) = \int_0^{\mu_1} d \mu \ e^\mu F(e^\mu)$$
where $\mu_1 = {\rm ln} \zeta_1$. This change of variable is very important not only for the adhesional
problems we consider here but also in other problems involving many length scales, e.g.,
rubber friction\cite{P8} or crack propagation\cite{PerssonBrener,PerssonUeba} in rubber. 

We now present an application relevant to the tire-rim sealing problem\cite{Cre}. 
A tire makes contact with the
wheel at the rim, where the rubber is squeezed against the steel with a pressure 
of order $0.5-1 \ {\rm MPa}$. The contact between
the rubber and the steel must be such that negligible air can leak through the non-contact
rubber-steel rim area. In addition, the 
rubber-rim friction must be so high that negligible slip occur even during strong
acceleration or breaking where the torque acting on the tire is highest. Since the rubber-rim friction 
originate from the area of real contact, this require a large fractional contact area $A(\zeta_1)/A_0$.

Consider an
elastic solid with a flat surface in adhesive contact 
with a hard, randomly rough
self affine fractal surface with the root-mean-square roughness 
(rms) $h_0= h_{\rm rms } =0.5 \ {\rm \mu m}$
and the fractal dimension $D_{\rm f} = 2.2$. 
The long and short distance cut-off wavevectors for the power spectrum $q_0=10^5 \ {\rm m^{-1}}$
and $q_1=10^{10} \ {\rm m^{-1}}$, respectively.
This gives a surface area $A^*$ which is about $38 \%$ larger than the projected
area $A_0$. 

The elastic solid has the (low-frequency) Young's modulus $E=10 \ {\rm MPa}$
and the Poisson ratio $0.5$,
as it typical for tire rubber.
The interfacial binding energy per unit area $\Delta \gamma =
0.05 \ {\rm J/m^2}$, as is typical for rubber in contact with
a steel surface. 

Fig \ref{A.gamm.hh0}(a) shows the relative contact area $A(\zeta_1)/A_0$ and (b) the (normalized) 
effective macroscopic interfacial energy per unit area
$\gamma_{\rm eff}(1)/\Delta \gamma$ as a function of $h/h_0$,
at vanishing applied pressure $\sigma_0 = 0$.
Here $h$ is the rms roughness used in the calculation, i.e., the actual power spectrum has
been scaled by the factor $(h/h_0)^2$.
We show results using both the ``old'' form of the elastic energy and the improved
form where $P(\zeta)$ is replaced by $\tilde P(\zeta)$. Note that the theory predict
$\gamma_{\rm eff}(1)=0$ for $h=h_0$, i.e., the theory predict that the macroscopic adhesion
vanish. 

In Fig. \ref{A.zeta} we show
(a) the relative contact area $A(\zeta)/A_0$ and (b) the (normalized) 
effective interfacial energy per unit area
$\gamma_{\rm eff} (\zeta)/\Delta \gamma$ as a function of magnification for $h=h_0$,
when the surfaces are squeezed together with the nominal pressure
$\sigma_0 = 0.5 \ {\rm MPa}$.
Note that the adhesion increases the area of real contact (at the highest magnification)
$A(\zeta_1)$
from $\approx 0.06 A_0$ to $\approx 0.44 A_0$. The adhesion becomes important already at
relative low magnification $\zeta \approx 10$ corresponding to the length scale
$\lambda = 2 \pi / (\zeta q_0) \approx 10 \ {\rm \mu m}$. 
The magnification at the point where adhesion becomes important can be estimated
by the condition that the nominal pressure $\sigma_0$ is of similar magnitude as the 
detachment stress: $\sigma_0 \approx (E \Delta \gamma q)^{1/2}$ which gives
$q \approx \sigma_0^2/E\Delta \gamma \approx 5\times 10^5 \ {\rm m}^{-1}$ or
$\zeta = q/q_0 \approx 5$.

\begin{figure}[htb]
   \includegraphics[width=0.45\textwidth]{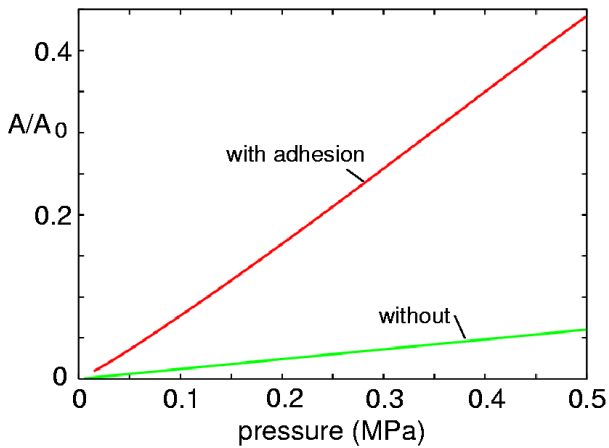}
\caption{
The relative contact area (at the highest magnification) $A(\zeta_1)/A_0$ 
as a function of the (nominal) squeezing pressure $\sigma_0$ with and without
the adhesional interaction.
For the same system as in
Fig. \ref{A.gamm.hh0}.
}
\label{A.P}
\end{figure}

Fig. \ref{A.P} shows the relative contact area (at the highest magnification) $A(\zeta_1)/A_0$ 
as a function of the (nominal) squeezing pressure $\sigma_0$ with and without
the adhesional interaction. For all squeezing pressures, the area of real contact is much larger when the
adhesional interaction is included.

Surface roughness is an important factor which influence the rate of leakage through seals.
If one assume that the contact regions at any magnification are approximately randomly 
distributed in the apparent contact area, we expect from percolation theory that the 
non-contact region will percolate when $A/A_0 \approx 1-p_{\rm c}$, where $p_{\rm c}\approx 0.7$ is the 
site percolation number. 
Since (because of adhesion)
at the highest (nanometer) resolution,
$A/A_0 \approx 0.44> 1-p_{\rm c} \approx 0.3$,
we conclude
that no roughness induced gas leakage will occur in the present case.

Next, let us focus on the tire-rim friction. The nominal contact area between
the tire rim and the steel is roughly $0.05 \ {\rm m^2}$. The kinetic frictional stress when 
rubber is sliding on a perfectly smooth clean 
substrate depend on the slip velocity but is typically of order\cite{RubberSmooth,Smooth1} 
$\approx 0.25 \ {\rm MPa}$. Thus, if the rubber is in perfect contact with the
steel, the shear force $\approx 1.3 \times 10^4 \ {\rm N}$. For the system studied above
we found $A/A_0 \approx 0.44$ giving the friction force $\approx 5.7 \ {\rm kN}$. If the load on a tire is
$3.5 \ {\rm kN}$ and the tire-road friction coefficient of order unity, the maximal friction force
$F_{\rm f} \approx 3.5 \ {\rm kN}$ which is 
below the estimated rubber-steel friction force, indicating negligible
slip of the tire relative to the steel. However, slip is sometimes observed, indicating that
the assumptions made above about, e.g., the steel surface roughness, may differ from real systems. 
It is clear, however, that analysis of the type presented above, using the measured surface topography
for the steel (or aluminum) surface, and the measured rubber viscoelastic modulus $E$ and interfacial
rubber-steel binding energy per unit area $\Delta \gamma$, may help in the design of systems where no
(or negligible) tire slip (and gas leak) occur at the rim.

\vskip 0.5 cm
{\bf 8.4 Detachment stress}

The expression (36) for the detachment stress will, in general, 
break down at very short length scales (high magnification).
This can be understood as follows.
For ``simple''  solids the elastic modulus $E$ can be related to the surface energy 
(per unit area) $\gamma_1 \approx Ea$, where $a$ is an atomic distance
(or lattice constant). Thus (36) predict the
that detachment stress at the highest (atomic) resolution $a$, where $\gamma_{\rm eff} = \Delta \gamma$ and
$q\sim 1/a$, is of order
$$\sigma_{\rm a} \approx (\Delta \gamma \gamma_1)^{1/2}/a$$ 
However, at the atomic scale the
detachment stress must be of order $\sigma^*=\Delta \gamma / a$.  
For many elastically stiff solids {\it with passivated surfaces} $\gamma_1 >> \Delta \gamma$. 
For such cases we suggest to use as an interpolation formula
$$\sigma_{\rm a} (\zeta) \approx { \sigma^*  \sigma^0_{\rm a} 
\over \sigma^* +\sigma^0_{\rm a}}$$
where
$$\sigma^0_{\rm a}= \left ({\alpha \gamma_{\rm eff} (\zeta) E 
q \over 1-\nu^2}\right )^{1/2}$$ 
We may define a cross-over length $l_{\rm c}$ via the condition\cite{PerssonWear}
$$\sigma^* = 
\left ({\gamma_{\rm eff} (\zeta) E 
\over l_{\rm c}}\right )^{1/2}$$
or
$$l_{\rm c} = 
{a^2 E 
\over \Delta \gamma}$$
For elastically stiff solids with passivated surfaces, 
the the cross-over length $l_{\rm c}$ may
be rather large, e.g., of order $l_{\rm c} \approx 1 \ {\rm \mu m}$ 
for silicon with the
surfaces passivated by grafted layers of chain molecules.
However, for elastically soft solids 
such as the PMMA adhering to most surfaces (see Sec. 10.1)
$\Delta \gamma \sim \gamma_1 \sim 0.1 \ {\rm J/m^2}$
so that the correction above becomes negligible. In this latter case, 
with $a\sim 1 \ {\rm \AA}$ the detachment stress at the highest
(atomic) magnification $\sigma_{\rm a} (\zeta_1) \approx 1 \ {\rm GPa}$. 
More generally, for unpassivated surfaces (even for stiff materials) 
$\gamma_1 \approx \Delta \gamma$ and the correction becomes unimportant.
However, most surfaces of hard solids have passivated surfaces under normal
atmospheric conditions and in these cases the ``correction'' 
above may become very important.

\vskip 0.5 cm
{\bf 8.5 Flaw tolerant adhesion}

The breaking of the adhesive contact between solids is usually due to the
growth of interfacial cracks which almost always exist due to the surface roughness
(see above) or due to other imperfections, e.g., trapped dust particles. Since the
stress at a crack tip may be much larger than the nominal (or average) stress at
the interface during pull-off, the adhesive bond will usually break at much lower nominal
pull-off stress than the ideal stress $\sigma^*=\sigma_{\rm a} (\zeta_1) 
\approx \Delta \gamma /a$, where $a$ is a 
typical bond distance of order Angstrom. An exception is the adhesion 
of very small solids (linear size $l < l_{\rm c}$),
where the pull-off force may approach the ideal 
value $\sigma^*$, see Sec. 8.4 and Ref. \cite{PerssonWear,Gao}. 

It has been suggested that the strength of adhesion 
for macroscopic solids can approach the ideal value
$\sigma^*$ if the material has an elastic modulus which increases as a 
function of the distance $z$ from
the surface (graded elasticity)\cite{Yao}. This can be understood as follows: 
The 
detachment stress for an interfacial crack of linear size $\lambda$ 
$$\sigma_{\rm a} \approx \left ({\gamma_{\rm eff} (\zeta) E 
\over \lambda(1-\nu^2)}\right )^{1/2}.\eqno(51)$$ 
Since the strain field of a crack of linear size $\lambda$ extend into the solid a distance
$\sim \lambda$, it will experience the elastic modulus for 
$z \approx \lambda$. Thus, the ratio $E/\lambda$ in (51) will be independent of the
magnification $\zeta \sim L/\lambda$ if
$E(z)\sim z$. Hence {\it if} $\gamma_{\rm eff}$ would be independent of the magnification 
$\zeta$, 
then the detachment stress $\sigma_{\rm a}(\zeta)$ would also be independent
of the magnification and of order the theoretical maximum $\sigma^*$. This argument was
presented by Yao and Gao\cite{Yao}. 
However, our calculations show that in most cases $\gamma_{\rm eff}$ will increase rapidly with increasing 
magnification, in which case the elastic modulus must increase faster than $z$ with the 
distance from the surface in order for $\sigma_{\rm a}$ to stay constant.  
In this case there will be no ``universal'' optimal form of the $z$-dependence of $E(z)$, 
since it will
depend on the surface roughness power spectrum.

\begin{figure}[htb]
   \includegraphics[width=0.3\textwidth]{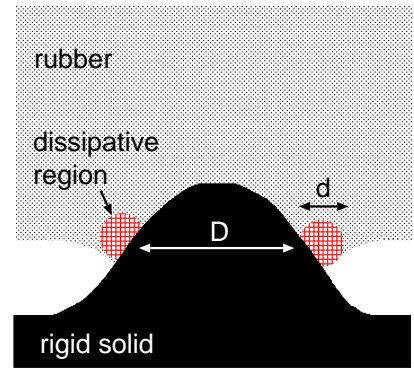}
\caption{
Asperity contact region during pull-off of a rubber block from a rough surface.
The dashed area indicate the dissipative region at the opening crack.
The theory of crack propagation in viscoelastic solids is only valid as
long as the linear size $d$ of the dissipative region at the crack tip
is smaller than the linear size $D$ of the asperity contact area.
}
\label{small}
\end{figure}

\vskip 0.5 cm
{\bf 8.6 Contact mechanics and adhesion for viscoelastic solids}

The contact between viscoelastic solids and hard, randomly rough, substrates 
has been studied in Ref. \cite{Hui} and
\cite{Cre} for stationary contact, and in Ref. \cite{P8} for sliding contact. 
In these papers the adhesional interaction between the solids was neglected. 
No complete theory has been developed for viscoelastic contact mechanics 
between randomly rough surfaces including the
adhesion. However, some general statements can be made. 

Assume that the loading or pull-off occur so slowly that the viscoelastic solid can be considered
as purely elastic everywhere except close to the asperity contact regions. 
That is, we assume that in the main part of the block the deformation rate is so slow
that the viscoelastic modulus $E(\omega) \approx E(0)$.
However, the detachment lines can be considered as interfacial cracks and close to the crack edges
the deformation rate may be so high and the full viscoelastic modulus must be used in these regions
of space.

The adhesional interaction is in general
much more important during separation of the solids than during the formation of contact (loading cycle).
This is related to the fact that separation involves the propagation of interfacial 
{\it opening} cracks, while the formation of contact (loading cycle) involves the propagation of 
interfacial
{\it closing} cracks. For flat surfaces, the 
effective interfacial energy (per unit area) necessary to propagate opening cracks is of
the form\cite{PerssonBrener,PerssonUeba,Greenw}
$$\gamma (v) = \Delta \gamma [1+f(v)]$$ 
where $f(v)$ is due to the viscoelastic energy dissipation in the vicinity of the
crack tip, and depend on the crack tip velocity $v$. For rubber-like viscoelastic solids,
$f(v)$ may increase with a factor of $\sim 10^3$ with increasing crack-tip velocity. Thus,
the viscoelastic deformations of the solid in the vicinity of the crack tip may strongly 
increase the pull-off force, as compared to the purely elastic case where $\gamma = \Delta \gamma$.
We note, however, that for a system where the contact only occur in small (randomly distributed)
asperity contact regions, the formula above will only hold as long as the 
linear size of the dissipative region is smaller
than the diameter of the contact area, see Fig. \ref{small}.

The effective interfacial energy gained during the propagation of closing cracks is of
the approximate form\cite{Greenw}
$$\gamma (v) \approx {\Delta \gamma \over 1+f(v)}$$ 
Thus, in this case $\gamma (v)$ is always smaller than (or equal to) $\Delta \gamma$,
so that the adhesional interaction is much less important than during pull-off.

\begin{figure}[htb]
   \includegraphics[width=0.47\textwidth]{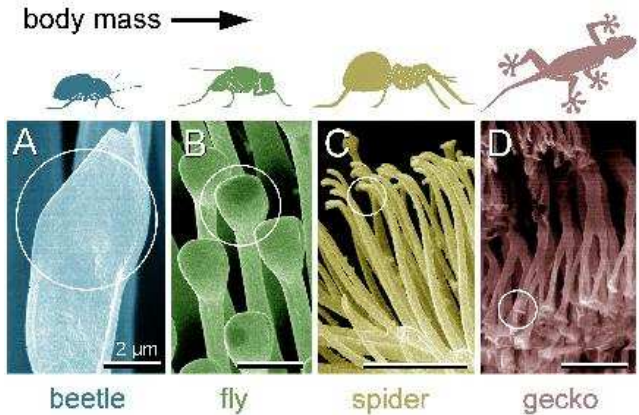}
\caption{
Terminal elements (circles) in animals with hairy design of attachment pads. Note that heavier 
animals exhibit finer adhesion structures.
Adapted from Ref. \cite{Arzt}.
}
\label{Gecko}
\end{figure}

\vskip 0.5 cm
{\bf 8.7 Strong adhesion for (elastically) stiff materials: Natures solution}

In order for an elastic solid to adhere to a rigid rough substrate, the elastic modulus of
the solid must be small enough. The physical reason is that 
the surface of the elastic solid must deform elastically for contact to
occur at the interface, and if the elastic energy necessary for the deformation is larger than the
interfacial binding energy $\Delta \gamma A_1$ (where $A_1$ is the area of real contact)
no adhesion will be observed.
Thus, experiment and theoretical studies have shown that already
a surface roughness with a root-mean-square amplitude of order a few micrometer is enough in order to
kill adhesion even for very soft materials such as rubber with a (low frequency) modulus of
order $\sim 1 \ {\rm MPa}$. In spite of this strong adhesion is possible even for intrinsically relative 
stiff materials
if instead of compact solids, the solids (or thick 
enough layers at the surfaces of the solids) form foam-like or fiber-like structures; 
the effective elastic properties of such non-compact solids can be much smaller
than for the compact solids. This fact is made use of in many biological adhesive systems.

The adhesive microstructures of lizards and many insects is the result of perhaps millions of 
years of development driven by the principle of natural selection. This has resulted in two different
strategies for strong adhesion: foam-like structures as for the adhesion pads of cicada (see Fig. \ref{Gorb}),
and fibers or hierarchical small-fibers-on-large-fibers structures 
(usually terminated by thin elastic plate-like structures), 
as for lizards, flies, bugs and
spiders, see Fig. \ref{Gecko}. These adhesive systems are made from keratin-like proteins with an elastic
modulus of order $\sim 1 \ {\rm GPa}$, i.e., $\sim 1000$ times higher than for rubber,  
but the effective elastic modulus of the non-compact solids can be very low\cite{Per}, 
making strong adhesion
possible even to very rough surfaces such as natural stone surfaces. 

\vskip 0.5 cm
{\bf 9 Elastoplastic contact mechanics with adhesion}

It is possible to apply the contact mechanics theory to the 
rather complex but very important\cite{YaPu} case of
elastoplastic contact with adhesion. In this case one need to solve the
equation for the stress distribution
$${\partial P \over \partial \zeta} = f(\zeta) {\partial^2 P \over \partial \sigma^2}$$
with the boundary conditions derived above
$$P(\sigma_{\rm Y},\zeta) = 0, \ \ \ \ \ \ \ \ \ \ 
P(-\sigma_{\rm a},\zeta) =0$$
and the ``initial'' condition
$$P(\sigma,1)=\delta (\sigma - \sigma_0)$$
If one introduce the new variable
$$s={\sigma_{\rm Y}-\sigma \over \sigma_{\rm Y}+\sigma_{\rm a}}$$
and consider $P(\sigma,\zeta)=\tilde P(s,\zeta)$ as a function of $s$ and $\zeta$,
and denote $\sigma_{\rm b} (\zeta)=\sigma_{\rm Y}+\sigma_{\rm a}(\zeta)$,
then the equations above takes the form
$${\partial \tilde P \over \partial \zeta} = s {\sigma'_{\rm b} \over \sigma_{\rm b}} 
{\partial \tilde P\over \partial s}
+{f(\zeta)\over \sigma_{\rm b}^2} {\partial^2 \tilde P \over \partial s^2}\eqno(52a)$$
$$\tilde P(0,\zeta) = 0, \ \ \ \ \ \ \ \ \ \
\tilde P(1,\zeta) =0\eqno(52b)$$
$$\tilde P(s,1)=\delta (\sigma_{\rm Y} - \sigma_0-s\sigma_{\rm b}(1))\eqno(52c)$$
The boundary conditions (52b) are satisfied if one expand
$$\tilde P(s,\zeta) =\sum_{n=1}^\infty A_n(\zeta) \ {\rm sin}(\pi n s)$$
Substituting this in (52a) gives a linear 
system of first order ordinary differential equations for $A_n(\zeta)$ which
can be integrated and solved using matrix inversion. We will present the detailed derivation and
numerical results elsewhere.

\vskip 0.5 cm
{\bf 10 Applications}

I present two applications involving very smooth surfaces, prepared by cooling
liquid glassy materials below the glass transition temperature (see Sec. 2). 
We first consider the contact between
a PMMA lens and a flat hard substrate, and discuss the 
experimental data of Bureau et al\cite{Bureau}.
We also consider contact mechanics for Pyrex glass surfaces, where plastic yield occur in
the asperity contact regions, and discuss the results of
Restagno et al\cite{Res}. 

\vskip 0.3 cm
{\bf 10.1 Contact mechanics for PMMA}

In the experiment by Bureau et al.\cite{Bureau}, a PMMA lens was prepared by cooling a liquid drop
of PMMA from $250 \ ^\circ {\rm C}$ to room temperature. In the liquid state the surface fluctuations of
vertical displacement are caused by thermally excited capillary waves (ripplons). When the liquid is near the
glass transition temperature $T_{\rm g}$ 
(about $100 \ ^\circ{\rm C}$ for PMMA), these fluctuations 
become very slow and are finally frozen in at the glass
transition. Thus, the temperature $T$ in (1) is 
not the temperature where the experiment was performed (room temperature), but rather
the glass transition temperature $T_{\rm g}$. 

\begin{figure}
  \includegraphics[width=0.45\textwidth]{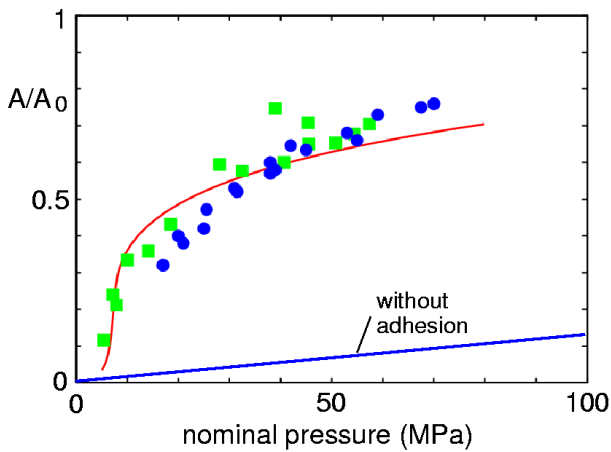}
  \caption{ \label{Baum}
The calculated relative contact area as a function of
the (nominal) pressure $\sigma_0$ including adhesion (upper solid line)
and without adhesion. 
Squares: the normalized, 
(nominal) shear stress $\tau /\tau_{
\rm f}$ for
PMMA sliding on TMS as a function of the (nominal) pressure.
The nominal shear stress $\tau$ has been divided by $\tau_{
\rm f} = 50 \ {\rm MPa}$,
which is the (measured) shear stress in the area of real contact (see text).
Circles: the normalized, 
(nominal) shear stress $\tau /\tau_{
\rm f}$ (with $\tau_{\rm f} = 5 \ {\rm MPa}$) for
PMMA sliding on OTS as a function of the (nominal) pressure.
}
\end{figure}

The (upper) solid line in Fig. \ref{Baum} shows the calculated relative contact area as a function of
the (nominal)  pressure $\sigma_0=F_{\rm N}/A_0$ (where $F_{\rm N}$ is the squeezing force). 
In the calculation we have used the equations above 
with the measured elastic (Young) modulus $E=2.9 \ {\rm GPa}$
and surface tension $\gamma = 0.04 \ {\rm J/m^2}$. 
We have also used the glass transition temperature $T_{\rm g} \approx 370 \ {\rm K}$,
the short-distance cut-off wavevector $q_1= 7\times 10^9 \ {\rm m}^{-1}$
and $q_c =1.7\times 10^9 \ {\rm m}^{-1}$ (see Eq. (1)). 
The cross-over wavevector $q_c$ 
(or the bending stiffness $\kappa$) has not been measured for
PMMA, but has been measured for other systems. Thus, for alkanes at $T\approx 100 \ ^\circ{\rm C}$
for C20 and C36, $q_c \approx 4.4\times 10^9$ and $\approx 2.7 \times 10^9 \ {\rm m}^{-1}$,
respectively\cite{Ocko}. There are also some studies of polymers\footnote{The
decrease as $\sim q^{-4}$ in the power spectral densities of 
polymers, as observed using the AFM 
\cite{Bollinne},  
start at a much smaller $q$ value than observed for small molar mass compounds.
A.M. Jonas (private communication) has suggested that this
could not be due to some kind of mechanical deformation of the
frozen capillary waves of smaller wavelength 
by the AFM tip. Longer wavelength waves would not be affected, whereas smaller ones 
(in the range of the radius of curvature of the apex of the AFM tip) 
would be dampened mechanically by the AFM tip, resulting in a lower than expected power density.}
\cite{Bollinne} (using the atomic force microscope)
giving $q_c \approx 1\times 10^8 \ {\rm m}^{-1}$. 
The interfacial binding energy $\Delta \gamma$ can be estimated 
using\cite{Is} $\Delta \gamma \approx 2 (\gamma_1 \gamma_2)^{1/2}
\approx 0.07 \ {\rm J/m^2}$, where $\gamma_1=\gamma \approx 0.04 \ {\rm J/m^2}$ is the surface
energy of PMMA and $\gamma_2 \approx 0.02 \ {\rm J/m^2}$ the surface energy of 
the (passivated) substrate. In the calculations we used $\Delta \gamma = 0.06 \ {\rm J/m^2}$.

If one assumes that, in the
relevant pressure range (see below), 
the frictional shear stress $\tau_{
\rm f}$ between PMMA and the substrate is essentially {\it independent} 
of the normal stress $\sigma$ 
in the asperity contact regions, 
then the calculated curve in Fig. \ref{Baum} 
is also the ratio between the nominal (or apparent) shear stress $\tau$ and the
true stress $\tau_{
\rm f}$; $A/A_0 = \tau /\tau_{\rm f}$ 
(note: the friction force $F_{\rm f} = \tau_{\rm f} A = \tau A_0$). 

The circles and squares in Fig. \ref{Baum} show the measured data of
Bureau et al.\cite{Bureau} for $\tau /\tau_{
\rm f}$. They performed
experiments where a PMMA lens, prepared by cooling a liquid drop of PMMA,
was slid on silicon wafers (which are nearly atomically smooth)
covered by a grafted silane layer. Two different types of 
alkylsilanes were employed for surface modification, namely a trimethylsilane (TMS)
and octadecyltrichlorosilane (OTS). The
squares in Fig. \ref{Baum} are the 
shear stress,
divided by $\tau_{
\rm f} = 50 \ {\rm MPa}$, for
PMMA sliding on TMS as a function of the (nominal) pressure $\sigma_0$.
The shear stress  $\tau_{
\rm f} = 50 \ {\rm MPa}$ was deduced from multi-contact 
experiments\cite{Bureau1} 
(where the local pressure in the asperity regions is so high as to give rise to local
plastic deformation) using
$\tau_{
\rm f} \approx \tau H/\sigma_0$, where $H \approx 300 \ {\rm MPa}$ is the hardness 
(yield stress) of PMMA
as determined from indentation experiments. 
This equation follows from $\tau A_0 = \tau_{\rm f} A$ and $A H = A_0 \sigma_0$.
The circles in Fig. \ref{Baum} show 
$\tau /\tau_{
\rm f}$ for
PMMA sliding on OTS as a function of the (nominal) pressure $\sigma_0$.
For this system no multicontact experiments were performed, and we have 
divided the apparent shear stress by  $\tau_{
\rm f}=5 \ {\rm MPa}$, chosen so as to
obtain the best agreement between the theory and experiment. Hence, in this case
$\tau_{
\rm f}$ represent a theoretically predicted shear stress for the multicontact
case; we suggest that multicontact experiments for the system PMMA/OTS 
are performed in order to test this prediction. 

The calculation presented above shows that the local pressure in the asperity
contact regions is of order $\sim 2 \ {\rm GPa}$, which is much higher than the 
macroscopic yield stress (about $300 \ {\rm MPa}$) of PMMA. 
However, nanoscale indentation experiments\cite{Japan} have shown
that on the length scale ($\sim 10 \ {\rm nm}$) and indentation depth scale ($\sim 1 \ {\rm nm}$)
which interest us here, the indentation hardness of PMMA is of order $\sim 10 \ {\rm GPa}$,
and no plastic yielding is expected to occur in our application.

The study above has assumed that the frictional shear stress at the
PMMA-substrate interface is independent of the local pressure $p=\sigma$,
which varies in the range $p \approx 50-100 \ {\rm MPa}$. This result is expected
because the shear stress will in general exhibit a negligible pressure
dependence as long as $p$ is much smaller than the {\it adhesional}
pressure $p_{\rm ad}$. The adhesional pressure is defined as follows: In order for
local slip to occur at the interface, the interfacial molecules must pass over
an energetic barrier of magnitude $\delta \epsilon$. At the same time the spacing
between the surfaces must locally increase with a small amount (some fraction of an
atomic distance) $\delta h$. This correspond to a pressure work $p a^2 \delta h$,
where $a^2$ is the contact area between the molecule, 
or molecular segment, and the substrate.
Thus the total barrier is of order $\delta \epsilon + p a^2 \delta h =
a^2 \delta h (p_{\rm ad}+p)$ where $p_{\rm ad} = \delta \epsilon / (a^2 \delta h)$.
For weakly interacting systems one typically 
have 
\footnote{If
one assumes that the energy necessary to move over the barrier $\delta \epsilon$ is lost during
each local slip event, then the shear stress $\sigma_{
\rm f} \approx (\delta \epsilon /a^2)/b$ where $b$
is the slip distance which will be of order a lattice constant. Thus, from the measured $\sigma_{
\rm f}$
(about $50 \ {\rm MPa}$ for TMS) and using $b\approx 3 \ {\rm \AA}$ gives $\delta \epsilon /a^2 \approx 
\sigma_{
\rm f} b \approx 1 \ {\rm meV/\AA^2}$ for TMS. For OTS $\delta \epsilon/a^2$ and the adhesion pressure
would be about 1 order of magnitude smaller than for TMS. This analysis does not take into account that 
the (measured) shear stress $\sigma_{
\rm f}$ involves {\it thermally-assisted} 
stress-induced dissipative events (i.e., $\sigma_{
\rm f}$ would be larger at lower temperatures),
so the estimation of $\delta \epsilon/a^2$ is a lower bound.} 
$\delta \epsilon /a^2 \approx 1 \ {\rm meV}/{\rm \AA}^2$
and $\delta h \approx 0.01 \ {\rm \AA}$ giving $p_{\rm ad} \approx 10 \ {\rm GPa}$, 
which is at least one order of magnitude larger than the pressures in the present experiment. 
We note that the pressure $p_{\rm ad}$ is similar to the  
pressure $p$ which must be applied 
before the viscosity of liquid hydrocarbon oils start to depend on the applied pressure.
The reason for the pressure dependence of the
viscosity of bulk liquids is similar to the problem above, involving the formation of
some local ``free volume'' (and the associated work against the applied pressure)
in order for the molecules to be able to rearrange during shear.

\begin{figure}
  \includegraphics[width=0.45\textwidth]{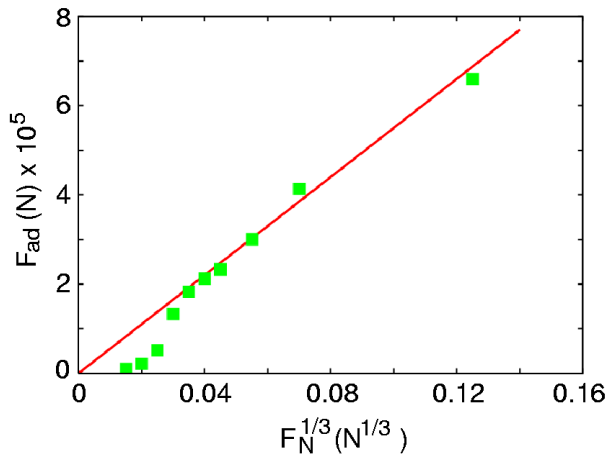}
  \caption{ \label{Data.Pyrex}
The adhesion force (or pull-off force) $F_{\rm ad}$ as a function
of the squeeze force $F_{\rm N}$ to the power $1/3$. The strait line is a
fit to the experimental data. Adapted from Ref. \cite{Res}
}
\end{figure}

\vskip 0.3 cm
{\bf 10.2 Contact mechanics for Pyrex glass}

Cohesion effects are of prime importance in powders and granular media, and they are
strongly affected by the roughness of the grain surfaces. For hydrophilic and elastically 
stiff solids such as silica glass, the cohesion is usually strongest under 
humid conditions because of the capillary forces exerted by small liquid bridges which form
at the contact between grains. Recently Restagno et al\cite{Res} have shown that
adhesive forces can be observed for stiff solids even in the absence of capillary bridges  
if the surfaces are smooth enough. 
However, even for very smooth Pyrex glass surfaces (see below) the adhesional force between a small
glass ball and a flat glass surface is
a factor $\sim 10^{-3}$ smaller than expected for {\it perfectly} smooth surfaces.
The elastic modulus of Pyrex glass is about $\sim 10^{2}$ times higher than for PMMA, 
and for this case the theory (see below)
predict that even an extremely small surface roughness will result in
{\it no adhesion}, assuming that only elastic deformation
occur in the contact areas between the Pyrex glass surfaces. However, the high contact pressure
in the asperity contact regions will result in local plastic yielding, and if the surfaces are squeezed
together with a high enough force, the plastically flattened regions will be so large
that the adhesional interaction from these contact regions
will overcome the (repulsive) force arising from elastic deformation
of the solids in the vicinity of the contact regions, resulting in a non-vanishing pull-off force.
This effect was observed in the study by Restagno et al.\cite{Res}.

In the experiments by Restagno et al.\cite{Res} a sphere with radius $R= 1.8 \ {\rm mm}$
was squeezed against a flat substrate, both of fire-polished Pyrex. The surface topography was studied using the AFM over an area $10\times 10 \ {\rm \mu m}^2$. The maximum height fluctuation was of order $\sim 1 \ {\rm nm}$
as expected from the frozen capillary waves. 
The adhesion experiments where performed in a liquid medium (dodecane) in order to avoid the
formation of capillary bridges.
If the ball and flat was squeezed together by a force
$F<F_c = 10^{-6} \ {\rm N}$ and then separated, no adhesion could be detected. 
However, if the squeeze force was larger
than this (critical) value, a finite pull-off force could be detected.

Fig. \ref{Data.Pyrex} shows the measured pull-off or adhesion force $F_{\rm ad}$ as a function
of $F_{\rm N}^{1/3}$, where $F_{\rm N}$ is the force with which the ball and the flat 
was squeezed together (i.e., the maximum force during the load cycle).
Note that $F_{\rm ad}$
depends approximately linearly on $F_{\rm N}^{1/3}$. As shown by Restagno et al.\cite{Res} this can be 
simply explained using standard contact mechanics for a ball-on-flat configuration, and assuming that
plastic yielding occur in all asperity contact regions.  

For elastically stiff solids such as Pyrex, 
the pull-off force for a perfectly smooth spherical object in adhesive contact with
a perfectly smooth flat substrate is given approximately by the Derjaguin approximation\cite{Derj}: 
$$F_{\rm ad} \approx 2 \pi \Delta \gamma R\eqno(53)$$
where $\Delta \gamma = 2 \gamma_{sl}$ is twice the interfacial energy (per unit area) of the Pyrex--dodecane
system. The same equation is valid when surface roughness occur, assuming that the dominant 
roughness has wavelength components smaller than the nominal contact area
between the solids. However, in this case $\gamma$ must be replaced with $\gamma_{\rm eff}$,
which is an effective interfacial energy which takes into account the surface roughness.  
The roughness modifies $\gamma$ in two ways. First, the area of real contact $A$ is smaller than
the nominal contact area $A_0$. This effect was taken into account by Restagno et al. by writing
the (macroscopic, i.e., $\zeta=1$)
$\gamma_{\rm eff} = \gamma A/A_0$. The second effect is the elastic energy stored in the vicinity of the 
contacting asperities. This elastic energy is given back during pull-off and will therefore
reduce $\gamma_{\rm eff}$. This effect was neglected by Restagno et al. 

During squeezing the nominal contact area is given (approximately) by Hertz's theory
$$A_0 = \pi \left ({FR \over E^*} \right )^{2/3}\eqno(54)$$
where $E^* =2E/3(1-\nu^2)$.
Restagno et al. assumed that plastic yielding has occurred in the area of real contact so that
$A=F/H$ where $H$ is the indentation hardness of the solids. Thus using (54):
$${A \over A_0} = {1\over \pi H} \left ({E^*\over R}\right )^{2/3}F^{1/3}\eqno(55)$$
Combining this with (53) and (55) gives
$$F_{\rm ad} = {4\gamma \over H} \left (R{E^*}^2\right )^{1/3} F^{1/3}\eqno(56)$$
The strait line in Fig. \ref{Data.Pyrex} is given by (56) 
with $\gamma=\gamma_{\rm sl} = 0.43 \ {\rm J/m^2}$, and with $R=1.8 \ {\rm mm}$, and the known elastic modulus, 
Poisson ratio and Pyrex hardness. 

\begin{figure}
  \includegraphics[width=0.45\textwidth]{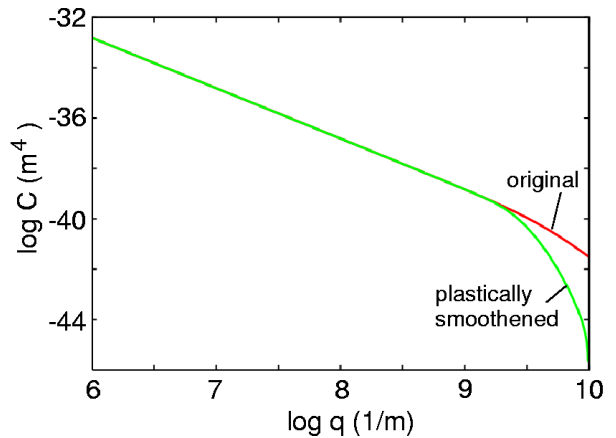}
  \caption{ \label{Pyrexa}
The surface roughness power spectrum of the original (upper curve) and the 
plastically smoothened surface.
For Pyrex glass using the power spectrum calculated
from the model with frozen capillary waves. In the calculation of the smoothened 
power spectrum we used the
elastic modulus $E=32 \ {\rm GPa}$ and Poisson 
ratio $\nu = 0.2$ and the hardness $\sigma_{\rm Y} = 6 \ {\rm GPa}$. 
}
\end{figure}

The theory above is consistent with the experimental data but does not address the following
fundamental questions:

(a) Why is no adhesion observed when the solids are brought together ``gently'', i.e., 
in the absence of the loading cycle (squeeze force $F_{\rm N}=0$)?

(b) Why does adhesion occur when the solids are first squeezed into contact. That is, why does not
the stored elastic energy at the interface break the interfacial bonds during 
removal of the squeezing force $F_{\rm N}$?

\begin{figure}
  \includegraphics[width=0.45\textwidth]{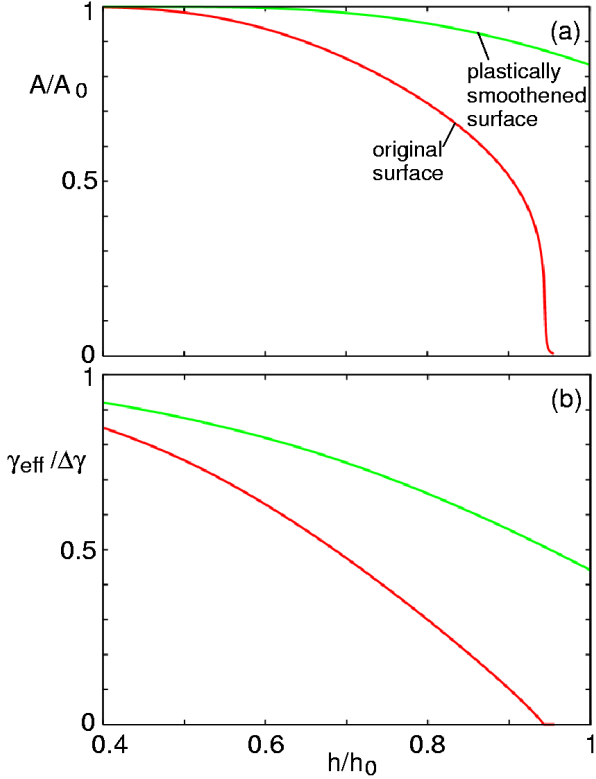}
  \caption{ \label{Pyrex2}
The calculated relative contact area $A/A_0$ (at the highest magnification)
and the effective interfacial binding energy
(per unit area) $\gamma_{\rm eff}$ as a function of
the surface roughness $h/h_0$. 
Here $h_0$ is the root-mean-square roughness of the actual surface and $h$ the rms 
roughness used in the calculation. Thus, the power spectrum has been multiplied with the factor
$(h/h_0)^2$.
(a) The contact area area
both for the plastically smoothened surface, and for the original surface. 
(b) The macroscopic interfacial energy $\gamma_{\rm eff}(1)$ in units of the bare interfacial
energy $\Delta \gamma$. 
For Pyrex glass using the power spectrum calculated
from the model with frozen capillary waves. In the calculation we used the
elastic modulus $E=32 \ {\rm GPa}$ and Poisson 
ratio $\nu = 0.2$ and $\Delta \gamma = 0.43 \ {\rm J/m^2}$. 
}
\end{figure}

We will now address these questions using the
contact mechanics theory for randomly rough surfaces described above. 
Assume that the surface roughness on
the Pyrex surfaces is due to frozen capillary waves. In fig. \ref{Pyrexa} 
(upper curve) we show the surface roughness power spectrum 
calculated using (1) with $T_{\rm g} =1100 \ {\rm K}$,
$\gamma = 1.0 \ {\rm J/m^2}$ and
$q_{\rm c}=0.5 \times 10^{\rm 10} \ {\rm m}^{-1}$. 
The lower curves in Fig. \ref{Pyrex2}(a) and (b) show
the relative contact area $A/A_0$ (at the highest magnification $\zeta=\zeta_1$)
and the effective macroscopic (i.e., for $\zeta=1$) interfacial binding energy
(per unit area) $\gamma_{\rm eff}(1)$ as a function of
the root-mean-square surface roughness $h$ divided by $h_0$. 
Here $h_0$ is the root-mean-square roughness of the actual surface and $h$ the rms 
roughness used in the calculation. Thus, the power spectrum has been 
multiplied (or scaled) with the factor
$(h/h_0)^2$. 
The macroscopic interfacial energy $\gamma_{\rm eff}(1)$ in (b) is in 
units of the bare interfacial
energy $\Delta \gamma$. 
Note that $\gamma_{\rm eff}(1)$ and the contact area both vanish for 
$h > 0.94 h_0$, so that no adhesion is expected in this case. 
This is in accordance with the experiment since
{\it no adhesion is observed if the glass surfaces are brought together without squeezing}.

\begin{figure}
  \includegraphics[width=0.45\textwidth]{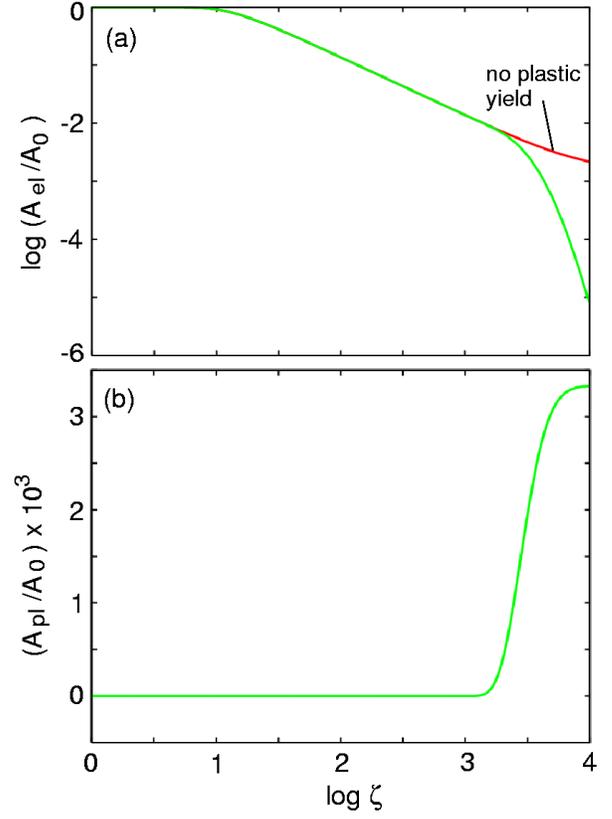}
  \caption{ \label{Pyrexb}
The calculated relative contact area as a function of
the magnification $\zeta$. (a) The elastic contact area 
both for the elasto-plastic model with the hardness 
$\sigma_{\rm Y}= 6 \ {\rm GPa}$ (lower curve), and 
for infinite hardness, where only elastic deformation occur (upper curve).
(b) The plastic contact area. For Pyrex glass using the power spectrum calculated
from the model with frozen capillary waves. In the calculation we used the
elastic modulus $E/2=32 \ {\rm GPa}$ 
(the factor $1/2$ arises from the fact that both solids are Pyrex with the same elastic
modulus $E$) and Poisson ratio $\nu = 0.2$. 
The magnification $\zeta = 10^4$ correspond to the
wave vector $q=\zeta q_0 =2\pi /\lambda = 10^{10} \ {\rm m^{-1}}$, i.e., to atomic resolution
$\lambda \approx 6 \ {\rm \AA}$. 
The plastic yielding occur in a relative narrow 
range of resolution; for $\zeta \approx 3000$, corresponding
to $\lambda \approx 2 \ {\rm nm}$, about half the contact area 
has yielded. Thus a plastic yielded solid-solid contact region have
typical area of order $4 \ {\rm nm}^2$. In the calculation the nominal
squeezing pressure $\sigma_0 = 20 \ {\rm MPa}$.
}
\end{figure}

Fig. \ref{Pyrexb} shows 
the calculated relative contact area as a function of
the magnification $\zeta$ using $C(q)$ 
from Fig. \ref{Pyrexa} as the surfaces are squeezed together with the nominal 
pressure $\sigma_0 = 20 \ {\rm MPa}$. 
In (a) we show the elastic contact area 
both for the elasto-plastic model with the hardness 
$\sigma_{\rm Y}= 6 \ {\rm GPa}$ (lower curve), and 
for infinite hardness, where only elastic deformation occur (upper curve).
In (b) we show the plastic contact area. 
Note that the magnification $\zeta = 10^4$ correspond to the
wave vector $q=\zeta q_0 =2\pi /\lambda = 10^{10} \ {\rm m^{-1}}$, i.e., to atomic resolution
$\lambda \approx 6 \ {\rm \AA}$. 
The plastic yielding occur in a relative narrow 
range of resolution; for $\zeta \approx 3000$, corresponding
to $\lambda \approx 2 \ {\rm nm}$, about half the contact area 
has yielded. Thus a plastic yielded solid-solid asperity contact region has
a typical area of order $4 \ {\rm nm}^2$. 
Note that in the experiment
(Fig. \ref{Data.Pyrex}) the nominal pressure $F_{\rm N}/A_0 = 
F_{\rm N}^{1/3} (E^*/R)^{2/3}/\pi$,
so the nominal pressure $\sigma_0 = 20 \ {\rm MPa}$ correspond 
to $F^{1/3}_{\rm N} = 0.06 \ {\rm N}^{1/3}$. 

\begin{figure}
  \includegraphics[width=0.3\textwidth]{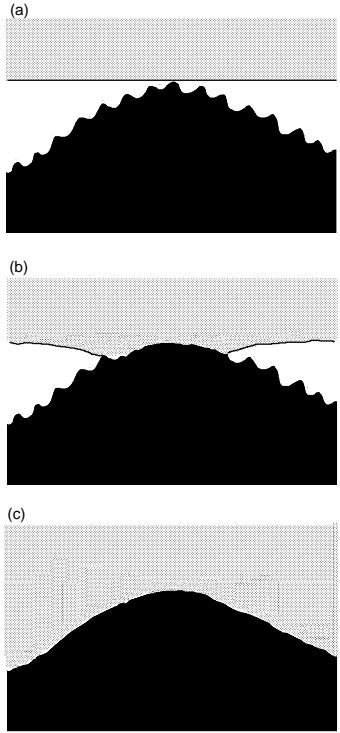}
  \caption{ \label{roughness.barrier}
An elastic solid (dotted area) in adhesive contact with a hard substrate (black) with
roughness on two different length scales.
(a) The (short-wavelength) 
substrate roughness is so high that no contact occur and the effective interfacial
energy $\gamma_{\rm eff}=0$.
(b) The small-wavelength roughness is flattened by plastic yield close to the top of the
large asperity by an earlier squeezing cycle (loading force $F_{\rm N}$). 
The contact area at zero load is determined 
by the adhesive interaction which is able to pull the surfaces into complete contact within the the 
smoothened region with an area $\sim F_{\rm N}$. 
The contact area cannot grow beyond the smoothened region because of the short
wavelength roughness outside the smoothened 
region--we refer to this as the 
{\it roughness barrier} effect. (c) In the model
calculation the surface is smoothened everywhere and the solids make contact nearly everywhere. 
}
\end{figure}

Fig. \ref{Pyrexb} shows that the short-ranged surface roughness will be plastically
flattened so that after squeezing the surfaces will conform at the shortest length-scale.
We can take this into account in the analysis of the pull-off force by replacing
the power spectrum $C(q)$ with a power spectrum $\bar C(q)$ corresponding to a ``smoothened''
surface, where the short wavelength roughness is removed. One approach would be to
cut-off $C(q)$ for $q>q_1^*$, where $q_1^*=q_0\zeta_1^*$  
where $\zeta_1^*$ is the magnification where, say $50\%$ of the contact area has yielded plastically.
Another approach is to define
$$\bar C(q_0\zeta) = \left (1-{A_{\rm pl}(\zeta)\over A_{\rm pl}^0} \right )C(q_0 \zeta)$$
where $A_{\rm pl}^0 = F_{\rm N}/\sigma_{\rm Y}$ is the contact area when the whole contact area has yielded
plastically. In Fig. \ref{Pyrexa} (lower curve) we show $\bar C(q)$ using
this definition. In Fig. \ref{Pyrex2} (a) and (b)
(upper curves) we show the
the relative contact area $A(\zeta_1)/A_0$ and the macroscopic effective interfacial energy
(per unit area) $\gamma_{\rm eff}(1)$ as a function of $h/h_0$ for the smoothened surface. 
Note that in this case for $h=h_0$
there is nearly perfect contact at the interface, and the macroscopic $\gamma_{\rm eff}$ is
about half of $\Delta \gamma$. In the actual applications the short-wavelength roughness 
is not removed (or smoothened) over the whole surface,
but only in small macro-asperity contact regions, with a total area proportional to the
maximum squeezing force $F_{\rm N}$ during the load cycle.
Nevertheless, the calculation shows that
one may expect nearly perfect contact within the smoothened area, and only a relative small reduction in
$\gamma_{\rm eff}$ due to the stored elastic energy at the interface. In fact, the reduction in
$\gamma_{\rm eff}(1)$ will be smaller than indicated in Fig. \ref{Pyrex2} because the elastic deformations will be smaller when
the solids only makes contact close to the (smoothened) macro-asperity 
tops (compare Fig. \ref{roughness.barrier} (b) and (c)).

In Fig. \ref{roughness.barrier} 
we illustrate (schematically) the basic results of our study. 
Here an elastic solid (dotted area) in adhesive contact with a hard substrate (black) with
roughness on two different length scales.
For the original surface (not plastically smoothened) (a) the (short-wavelength) 
substrate roughness is so high that no contact occur and the effective interfacial
energy $\gamma_{\rm eff}=0$. During the loading cycle, the short wavelength roughness 
at the top of the macro-asperities
is flattened by plastic yield.
For the smoothened surface the contact area at zero load is determined 
by the adhesive interaction which is able to pull the surfaces into complete contact within the the 
smoothened region (see (b)), with an area proportional to the maximum loading force $F_{\rm N}$ during the
loading cycle. 
The contact area cannot grow beyond the smoothened region because of the short
wavelength roughness outside the smoothened 
region--we refer to this as the 
{\it roughness barrier} effect. 
In the model
calculation the surface is smoothened everywhere as in (c) and 
the solids makes contact nearly everywhere. 

\vskip 0.5 cm
{\bf 11 On the philosophy of contact mechanics}

\begin{figure}
  \includegraphics[width=0.3\textwidth]{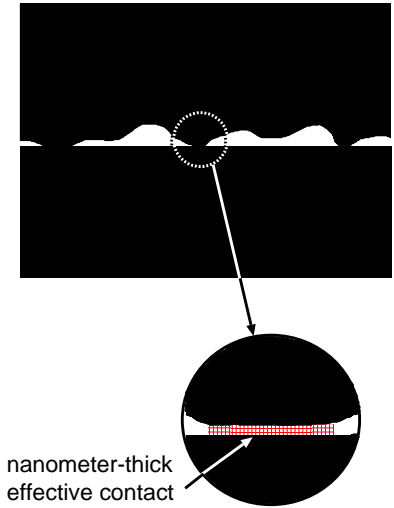}
  \caption{ \label{nano}
A fundamental approach to contact mechanics, adhesion and friction 
consist of applying continuum
mechanics down to a length scale of order a few nanometer. At this length scale or resolution continuum 
mechanics will correctly predict the contact area and the pressure distribution. At shorter length scale
other methods such as molecular dynamics or quantum mechanical 
methods must be employed which takes into account the atomistic and chemical nature of real surfaces.
}
\end{figure}

Recently it has been argued that contact mechanics models based on continuum mechanics
cannot be used to determine the interfacial pressure distribution
with atomistic resolution, or the adhesion or friction between the contacting solids,
since these properties depend
on atomic scale surface roughness (surface adatoms, steps, and so on)\cite{RobbinsM}. 
But the aim of continuum mechanics was never to describe the physics at the atomistic level but only
at larger length scales. 
Thus, a fundamental approach to the 
problems mentioned above consist of applying continuum
mechanics down to a length scale of order a few nanometer. At this length scale or resolution continuum 
mechanics will correctly predict the contact area and the pressure distribution. At shorter length scale
other methods such as molecular dynamics or quantum mechanical 
methods must be employed which takes into account the atomistic and chemical nature of real surfaces.
However, in most practical applications such a detailed approach will not be possible because
too little is usually known about the actual interface, e.g., thin (nanometer or less) 
contamination layers may
dominate the physics at the atomistic level. Nevertheless, even in these cases the continuum
mechanics approach may be extremely useful as it determines the size and the pressure
distribution in the {\it effective} contact
areas where most of the interesting physics takes place, see Fig. \ref{nano}.

\vskip 0.5 cm
{\bf 12 Comment on numerical studies of contact mechanics}

Many numerical studies based on the finite element method, or other numerical methods,
have been presented for the elastic or elastoplastic contact between randomly
rough (e.g., self affine fractal) surfaces\cite{summera1,summera2,summera3,summera4}. 
However, numerical methods cannot be used to study
the influence of roughness on the contact mechanics for macroscopic systems involving
perhaps 7 decades in length scales (from nanometer to cm). 
The computational time, and the required memory space, scales at least linearly 
with the number of grid points $N=(L/a)^2$ in the $xy$-plane 
($a$ is the grid size and $L$ the linear size of the
system studied). (The linear scaling $\sim N$ assumes using a 
multi-scale approach\cite{YTP}.) 
Thus, increasing the linear
size of the system with one decade $L\rightarrow 10L$ will, at 
least, increase the needed memory size and 
the computational time by a factor of 100, 
and at present converged (see below) numerical studies 
are limited to systems with 2 or maximum 3 decades in length scale.
Nevertheless, numerical studies even on relative 
small systems may be useful to test the accuracy of
analytical contact mechanics theories.

\begin{figure}
  \includegraphics[width=0.45\textwidth]{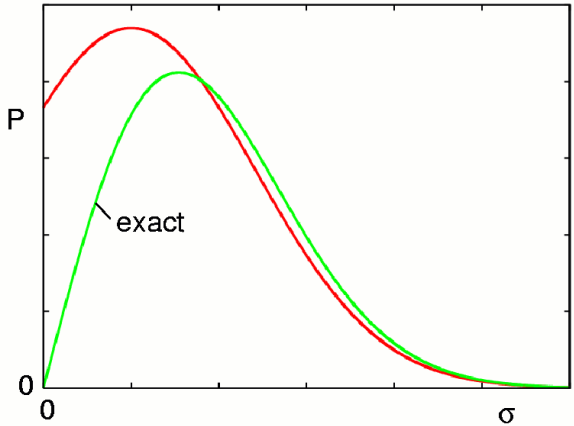}
  \caption{ \label{Gauss}
The exact and approximate stress distribution $P(\sigma, \zeta)$.
The exact $P(\sigma,\zeta)$ vanish for $\sigma=0$, but this result is not observed
in numerical calculations owing to a too sparse integration grid (see text for details). 
The two stress distribution functions correspond to the same load, but the approximate stress
distribution gives $22\%$ larger area of contact.
}
\end{figure}

All numerical studies of contact mechanics for randomly rough surfaces I am aware of
are {\it non-converged} in the following sense. Assume that the wavelength of the shortest 
surface roughness component
is $\lambda_1$. In order to properly describe contact mechanics it is necessary to have at 
least $10\times 10$ grid points and preferably much more 
(maybe $100\times 100$) within one asperity contact region 
which may be smaller than, or of order $ \lambda_1\times \lambda_1$.
Only when this smallest Hertzian-like contact problem is accurately described can one obtain accurate
results for the area of real contact. It is easy to show that when too few grid points are used, the
numerical study will give an {\it overestimation} of the area of real contact. The reason for this is that
the true stress distribution vanishes at zero stress 
(see Sec. 4), and in order to obtain
this result (approximately) in numerical studies it is necessary to have such a high density of grid points that
the contact region close to the non-contact area is accurately described (it is this region which determine the
boundary condition $P(0,\zeta)=0$). In all reported studies I am aware of this is not the case and the 
stress distribution remain finite (and large) at $\sigma =0$ as indicated in Fig. \ref{Gauss}. 
However, since the
applied load $F_{\rm N}$ is determined by the first moment of the pressure distribution,
$$F_{\rm N}=A_0\int_0^\infty d\sigma \ \sigma P(\sigma,\zeta)$$
it follows that the numerically determined $P(\sigma, \zeta)$ curve must be below the exact 
stress distribution curve for larger $\sigma$ in order for $F_{\rm N}$ to be the same. 
But since the area of real contact $A$ is given by
$$A= A_0 \int_0^\infty d\sigma \ P(\sigma,\zeta)$$
it follows that the numerically determined contact area will be larger than the actual one.

\vskip 0.5 cm
{\bf 13 Summary and conclusion}

I have reviewed basic contact mechanics theories. For surfaces with random roughness on
many different length scales, the contact area (in the absence of adhesion)
is proportional to the squeezing force at low
enough squeezing force.
In addition, theory and numerical simulations show that 
when the contact area is proportional to the squeezing force,
the stress distribution at the interface is
independent of the squeezing force and the distribution
of sizes of the contact regions does not depend on the squeezing force.
Thus, when the squeezing
force increases, new contact regions are formed in such a way that the distribution of
contact regions and the pressure distribution remains unchanged. This is the physical origin
of Coulomb's friction law which states that the friction force is proportional to the
normal (or squeezing) force\cite{BookP}, and which usually holds accurately as long as the block-substrate
adhesional interaction can be neglected.

I have presented a detail discussion about the boundary conditions obeyed by
the stress probability distribution $P(\sigma, \zeta)$ 
for elastic, elastoplastic and adhesive contact between solids.
These boundary conditions are needed when studying contact mechanics using the formalism
developed in Ref. \cite{P8}.
I have presented numerical results illustrating some aspects of the theory.
I have analyzed contact mechanics problems for very smooth polymer
(PMMA) and Pyrex glass surfaces prepared by
cooling liquids of glassy materials from above the glass transition temperature,
and shown that the surface roughness which results from capillary waves 
when a glassy material
is cooled below the glass transition temperature
can have a large influence on the contact mechanics between the solids. 
The analysis suggest a new explanation for 
puzzling experimental results [L. Bureau, T. Baumberger and C. Caroli,
arXiv:cond-mat/0510232] about the dependence
of the frictional shear stress on the load for contact between a glassy polymer 
lens and flat substrates. I have also studied the contact mechanics between Pyrex glass surfaces
where plastic deformation occur at the nanometer length scale. Finally, I have given some comments about
the possibility of testing analytical contact mechanics theories using numerical methods, e.g., fine element
calculations. 

\vskip 0.3cm
{\bf Acknowledgement}
I thank T. Baumberger and L. Bureau for drawing my attention to their interesting experimental
data and for supplying me the data used in Fig. \ref{Baum}, and E. Charlaix for drawing my attention to 
Ref. \cite{Res}. 
I thank Pirelli Pneumatici for support.

\vskip 0.5cm
{\bf Appendix A: Power spectrum of frozen capillary waves}

Surfaces of liquids have thermally excited capillary waves. 
There are three contributions to the
potential energy of capillary waves: a gravitational contribution, a surface tension contribution and
another contribution which depend on the curvature of the liquid surface. The surface tension
contribution depend on the surface area and can be written as
$$\gamma \int d^2x \left (1+ (\nabla h )^2\right )^{1/2} \approx
\gamma \int d^2x \left (1+ {1\over 2} (\nabla h )^2\right )$$
$$= \gamma A_0 + {1\over 2}\gamma \int d^2x (\nabla h )^2 $$
The gravitational contribution is of the form
$${1\over 2} \int d^2x \ \rho g h^2$$
Finally the curvature contribution to the potential energy is
$$ {1\over 2} \kappa \int d^2x \left (\nabla^2 h \right )^2
$$
Thus the total potential energy
$$U=
{1\over 2} \int_{A_0} d^2x \left ( \rho g h^2
+\gamma (\nabla h )^2 
+\kappa  \left (\nabla^2 h \right )^2 \right )$$
If we write
$$h({\bf x}) = \sum_{\bf q} h({\bf q}) e^{i{\bf q}\cdot {\bf x}}\eqno(A1)$$
we get
$$U=
{1 \over 2} A_0 \sum_{\bf q} \left ( \rho g
+\gamma q^2 
+\kappa  q^4 \right )|h({\bf q})|^2$$
Each term in this expression is similar to the potential 
energy of a harmonic oscillator which on the average
(in the classical limit) must equal $k_{\rm B}T/2$ so that
$$A_0 \left ( \rho g
+\gamma q^2 
+\kappa  q^4 \right )\langle |h({\bf q})|^2\rangle =k_BT\eqno(A2)$$
The power spectrum can be written as\cite{P8}
$$C(q)={1\over (2\pi )^2}\int d^2x \ \langle h({\bf x})h({\bf 0})e^{-i{\bf q}\cdot {\bf x}}\eqno(A3)$$
Substituting (A1) in (A3) gives
$$C(q)= {A_0 \over (2\pi )^2} \langle |h({\bf q})|^2\rangle \eqno(A4)$$
Combining (A2) and (A4) gives
$$C(q)= {1\over (2 \pi)^2} 
{k_{\rm B} T \over \rho g
+\gamma q^2 
+\kappa  q^4 }$$

Since the potential energy is a quadratic function of the dynamical variable $h({\bf q})$,
the distribution of heights $P_h = \langle \delta (h-h({\bf x}))\rangle$ will be Gaussian.
The difference $\Delta h$ between the highest and lowest point on a
randomly rough surface with a Gaussian height distribution, 
$$P(h) = (2 \pi h_{\rm rms}^2)^{-1/2}
{\rm exp} (-h^2/2h_{\rm rms}^2),$$ 
depend on the size (area $A_0$) of the surface, 
but is in general much higher
than the root-mean-square amplitude $h_{\rm rms}$. 
For example, if the height correlation function
$C(|{\bf x}|)=\langle h({\bf x})h({\bf 0})\rangle $ has a characteristic decay length $\xi$
(i.e., $C(\xi )/C(0) << 1$) then there will be roughly $N=A_0/\xi^2$ uncorrelated 
points on the surface.
Thus $\Delta h \approx 2h_1$ where $h_1$ is obtained from 
$\int_{h_1}^\infty dh P(h) \approx 1/N$. 
For $N >> 1$ this gives
$${h_1 \over h_{\rm rms}} =  \left [ 2  {\rm ln} \left ({ h_{\rm rms} N 
\over h_1 \surd (2\pi) }\right ) \right ]^{1/2}$$
For example, if $A=1 \ {\rm cm}^2$ and
$\xi = 1 {\rm \mu m}$ we get $N=10^8$ and 
$h_1 \approx 6 h_{\rm rms}$ and thus 
$\Delta h \approx 12 h_{\rm rms}$.
If instead $N=10^4$ we get 
$\Delta h \approx 8 h_{\rm rms}$, so that $\Delta h$ is not very sensible to the exact value of
$N$ and in general $\Delta h \approx 10 h_{\rm rms}$.
Restagno et al studied a $10 \ {\rm \mu m}\times 10
{\rm \mu m}$ Pyrex glass surface region using AFM ($N=512\times 512$ data points) and
observed a hight fluctuation to be about $2 \ {\rm nm}$. This is about 10 times 
higher than the root-mean-square height fluctuation $h_{\rm rms}$ expected for frozen 
capillary waves, and is
roughly what is expected theoretically in this case.

\vskip 0.5cm
{\bf Appendix B: Stress distribution in adhesive contact between smooth spherical surfaces}

\begin{figure}
  \includegraphics[width=0.45\textwidth]{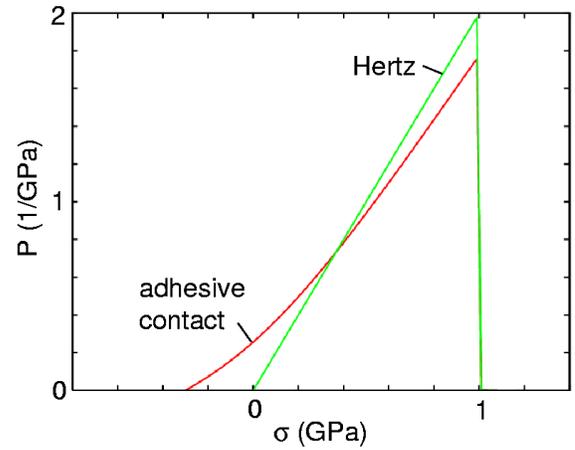}
  \caption{ \label{Pdist}
The stress distribution $P(\sigma )$ in the area of contact between two
elastic spheres in adhesive contact and without adhesion (Hertz case).
In the calculation we have assumed $\sigma_{\rm max}=1 \ {\rm GPa}$ and the
adhesive detachment stress $\sigma_{\rm a}= -0.3 \ {\rm GPa}$, and the parameter
$\eta=1.1$.
}
\end{figure}

Greenwood and Johnson\cite{GJ} have presented a very simple and useful approach to the contact mechanics between
solids with locally spherical (or quadratical) surfaces. 
They observed that the {\it difference}
between two Hertzian pressure distributions, corresponding to two different values $a$ and $b$ 
of the radius of
the circular contact area, describes adhesive contact between the solids. 
The stress for $r<a$ 
$$\sigma = A\left (a^2- r^2\right )^{1/2}
-B\left (b^2- r^2\right )^{1/2}$$ 
where $A>B>0$ and $b>a$,
while for $a<r<b$
$$\sigma = 
-B\left (b^2- r^2\right )^{1/2}$$ 
so that $\sigma$ is continuous for $r=a$. In this model the spheres are in contact for 
$r<a$. For $r>a$ no contact occur but there is an attractive interaction 
between the walls for $a<r<b$, while for $r>b$ no forces act between the walls.
Thus, we can consider the circular strip $a<r<b$ as the crack tip process zone. 
Since the stress is finite everywhere on the solid surfaces, a nonzero crack process zone is necessary
in order for the work of adhesion $\gamma$ to be nonzero.
If we introduce $\eta = b/a$ and denote the maximum compressional stress (which occur for $r=0$) 
as $\sigma_{\rm max}$, and
the maximal tensal stress (which occur for $r=a$) with $\sigma_{\rm a}$, and if we measure
$r$ in units of $a$ we get for $0 < r < 1$
$$\sigma = \left [\sigma_{\rm max}+{\sigma_{\rm a} \eta \over (\eta^2-1)^{1/2}}\right ] (1-r^2)^{1/2}$$
$$-{\sigma_{\rm a} \over (\eta^2-1)^{1/2}} (\eta^2-r^2)^{1/2}$$
while for $1 < r <\eta$
$$\sigma = -
{\sigma_{\rm a} \over (\eta^2-1)^{1/2}} (\eta^2-r^2)^{1/2}$$
The applied normal force or load is given by
$$F_{\rm N} = \int_{r\leq b} d^2x \ \sigma({\bf x})= 
{2\pi \over 3} 
\left (\sigma_{\rm max}-\eta (\eta^2-1)^{1/2}\sigma_{\rm a}\right )$$
and the normal force acting in the contact area
$$F_0=
\int_{r\leq a} d^2x \ \sigma({\bf x})= 
F_{\rm N}+{2\pi \over 3}\sigma_{\rm a} (\eta^2-1)$$
Note that $F_0>F_{\rm N}$ as it must because the normal force in the contact area must equal
the external load {\it plus} the additional ``adhesion load'' 
coming from the attraction between the solids in the 
non-contacting ($r>a$) region.
The pressure distribution in the contact area
$$P(\sigma) = {1\over \pi a^2}\int_{r\leq a} d^2x \ \delta (\sigma -\sigma({\bf x}))$$
This function is shown in Fig. \ref{Pdist} for $\sigma_{\rm max}=1\ {\rm GPa}$, 
$\sigma_{\rm a}=0.3 \ {\rm GPa}$
and $\eta=1.1$ corresponding to a crack tip process zone of width $b-a = 0.1a$. 
Note that $P(\sigma)$ vanish continuously at $\sigma = - \sigma_{\rm a}$.
In fact, it is easy to show that close to the detachment stress
$\sigma=-\sigma_{\rm a}$, 
$$P(\sigma)\approx {2(\sigma+\sigma_{\rm a}) \over \sigma_1^2}$$
where
$$\sigma_1 = \sigma_{\rm max} +{{\sigma_{\rm a}}\eta \over (\eta^2-1)^{1/2}}$$
For $\sigma_{\rm a}=0$, $\sigma_1 = \sigma_{\rm max}$ 
we obtain the Hertz solution, Eq. (23).

\begin{thebibliography}{999}


\bibitem{Hertz1}
H. Hertz,
J. reine und angewandte Mathematik {\bf 92}, 156 (1882).

\bibitem{BookP}
See, e.g., B.N.J. Persson, {\it Sliding Friction: Physical Principles and Applications},
Second ed., Springer, Heidelberg, 2000. 

\bibitem{Archard}
J.F. Archard, 
Proc. Roy. Soc. London, Ser. A{\bf 243}, 190 (1957).

\bibitem{GreenW} J.A. Greenwood and J.B.P. Williamson, Proc. Roy. Soc. London,
Ser. A{\bf 295}, 300 (1966). 

\bibitem{Fuller}
K.N.G. Fuller and D. Tabor, Proc. R. Soc. London, Ser. A{\bf 345}, 327 (1975).

\bibitem{Bush} A.W. Bush, R.D. Gibson and T.R. Thomas, Wear {\bf 35}, 87 (1975);
A.W. Bush, R.D. Gibson and G.P. Keogh, Mech. Res. Commun. {\bf 3}, 169 (1976).

\bibitem{GZ} 
G. Zavarise, M. Borri-Brunetto and M. Paggi, WEAR (to be published 2006). 


\bibitem{P66} B.N.J. Persson, F. Bucher, B. Chiaia 
                     Phys. Rev. B{\bf 65}, 184106 (2002)

\bibitem{P8} 
B.N.J. Persson, J. Chem. Phys. {\bf 115}, 3840 (2001).

\bibitem{P1}
B.N.J. Persson, Eur. Phys. J. E{\bf 8}, 385 (2002).

\bibitem{Cre}
B.N.J. Persson, O. Albohr, C. Creton and V. Peveri, J. Chem. Phys. {\bf 120}, 8779 (2004).


\bibitem{P3}
B.N.J. Persson, O. Albohr, U. Tartaglino, A.I. Volokitin and E. Tosatti, 
J. Phys. Condens. Matter {\bf 17}, R1 (2005) 

\bibitem{Nayak}
P.R. Nayak, ASME J. Lubr. Technol. {\bf 93}, 398 (1971).

\bibitem{Lag}
F. Laguarta, N. Lupon and J. Armengol, Applied Optics {\bf 33}, 6508 (1994).

\bibitem{Jack}
J. J\"ackle and K. Kawaski,
J. Phys.: Condens. Matter {\bf 7}, 4351 (1995).

\bibitem{Die}
K.R. Mecke and S. Dietrich,
Phys. Rev. E{\bf 59}, 6766 (1999)

\bibitem{Buzza}
D.M.A. Buzza, Langmuir {\bf 18}, 8418 (2002).

\bibitem{Sey}
T. Seydel, M. Tolan, B.M. Ocko, O.H. Seeck, R. Weber, E. DiMasi, and W. Press,
Phys. Rev. B{\bf65}, 184207 (2002).

\bibitem{JKR}
K.L Johnson, K. Kendall and A.D. Roberts, Proc. R. Soc. London, 
Ser. A{\bf 324}, 301 (1971).

\bibitem{Is}
J.N. Israelachvili, {\it Intermolecular and Surface Forces} (London: Academic, 1995).

\bibitem{Derj}
B.V. Derjaguin, Kolloid Zeits. {\bf 69}, 155 (1934). See also Ref. \cite{Is}.


\bibitem{Derja}
B.V. Derjaguin, Wear {\bf 128}, 19 (1988); D. Tabor, in {\it Surface Physics of Materials}, Vol. II,
edited by J.M. Blakely (Academic University Press, New York, 1975) p. 475.

\bibitem{Flash}
B.N.J. Persson, to be published.

\bibitem{Jon}
K.L. Johnson, {\it Contact Mechanics} (Cambridge University Press,
Cambridge, 1985).

\bibitem{summera4}
S. Hyun, L. Pei, J.F. Molinari and M.O. Robbins, Phys.
Rev. {\bf E}70, 026117 (2004).


\bibitem{foam}
A foam-like structure has an effective Young's modulus proportional to
$(d/D)^3$, where $D$ is the cell size and $d$ the wall thickness, see 
L.J. Gibson and M.F. Ashby, Proc. Roy. Soc. Lond. A{\bf 382}, 43 (1982).

\bibitem{Sherg}
M. Sherge and S. Gorb, {\it Biological micro- and nano-tribology--Nature's solutions}
(Springer, Berlin, 2001).

\bibitem{LizardP}
B.N.J. Persson and S. Gorb, J. Chem. Phys. {\bf 119}, 11437 (2003).

\bibitem{LizardCP}
G. Carbone, L. Mangialardi and B.N.J. Persson, Phys. Rev. B{\bf 70}, 125407 (2004).

\bibitem{GJ}
J.A. Greenwood and K.L. Johnson, J. Phys. D{\bf 31}, 3279 (1998).

\bibitem{PerssonBrener}
B.N.J. Persson and E. Brener, Phys. Rev. E{\bf 71}, 036123 (2005);
C. Carbone and B.N.J. Persson, Eur. Phys. J. E{\bf 17}, 261 (2005).

\bibitem{PerssonUeba}
B.N.J. Persson, O. Albohr, G. Heinrich and H. Ueba,
J. Phys.: Condens. Matter {\bf 17}, R1071 (2005).

\bibitem{RubberSmooth}
K. Vorvolakos and M.K. Chaudhury, Langmuir {\bf 19}, 6778 (2003).

\bibitem{Smooth1}
M. Barquins and A.D. Roberts, J. Phys. D: Appl. Phys. {\bf 19}, 547 (1986).

\bibitem{PerssonWear}
B.N.J. Persson, WEAR {\bf 254}, 832 (2003).

\bibitem{Gao}
H. Gao and H. Yao, Proc. Natl. Acad. Sci. USA {\bf 101}, 7851 (2004).

\bibitem{Yao}
H. Yao and H. Gao, 
J. Mechanics and Physics of Solids, in press (2006). 

\bibitem{Hui}
C.Y. Hui, Y.Y. Lin, and J.M. Barney, J. Polym. Sci., Part B: Polym. Phys. {\bf 38}, 1485 (2000).

\bibitem{Greenw}
J.A. Greenwood, J. Phys. D: Appl. Phys. {\bf 37}, 2557 (2004).

\bibitem{Per}
B.N.J. Persson, J. Chem. Phys. {\bf 118}, 7614 (2003).

\bibitem{Arzt}
E. Arzt, S. Gorb and R. Spolenak,
PNAS {\bf 100}, 10603 (2003).

\bibitem{YaPu}
L. Zhang and Ya-Pu Zhao,
J. Adhesion Sci. Technol. {\bf 18}, 715 (2004).

\bibitem{Bureau}
L. Bureau, T. Baumberger and C. Caroli,
arXiv:cond-mat/0510232 v1.

\bibitem{Res}
F. Restagno, J. Crassous, C. Cottin-Bizonne and E. Charlaix, Phys. Rev. E{\bf 65}, 042301 (2002).

\bibitem{Ocko}
B.M. Ocko, X.Z. Wu, E.B. Sirota, S.K. Sinha, and M. Deutsch,
Phys. Rev. Lett. {\bf 72}, 242 (1994).

\bibitem{Bollinne}
C. Bollinne, S. Cuenot, B. Nysten and A.M. Jonas,
Eur. Phys. J. E{\bf 12}, 389 (2003).

\bibitem{Bureau1}
L. Bureau, T. Baumberger and C. Caroli,
Eur. Phys. J. E{\bf 8}, 331 (2002).

\bibitem{Japan}
See T. Ito, Y. Taguchi and T. Uematsu, (ELIONIX Co. Ltd. Tokyo, Japan),  
home.elionix.co.jp/dataphto/jmtra02.pdf

\bibitem{RobbinsM}
B.Q. Luan and M.O. Robbins,
NATURE {\bf 435}, 929 (2005).

\bibitem{summera1}
M. Borri-Brunetto, A. Carpinteri and B. Chiaia, International Journal of Fracture {\bf 95},
221 (1999).

\bibitem{summera2}
M. Borri-Brunetto, B. Chiaia and M. Ciavarella, Comput. Methods Appl. 
Mech. Eng. {\bf 190}, 6053 (2001).
 
\bibitem{summera3}
A. H\"onig, {\it Abschlu\ss bericht f\"ur das Project: Kontact- und
Reibungsverhalten von amorphen Schichten} (unpublished) (2004). 

\bibitem{YTP}
C. Yang, U. Tartaglino and B.N.J. Persson, Eur. Phys. J. E{\bf 19}, 47 (2006).

\end{thebibliography}
\end{document}